\documentclass[aps,pra,twocolumn,showpacs,superscriptaddress,groupedaddress]{revtex4-2}
\usepackage[utf8]{inputenc}
\usepackage{amsmath, amsthm, amssymb, amsfonts,enumitem,listings}
\usepackage[makeroom]{cancel}
\usepackage{graphicx}
\usepackage{bm}
\usepackage{dsfont} 
\usepackage{mathtools}
\usepackage{booktabs}
\usepackage{thm-restate}
\usepackage{comment}
\usepackage{amsmath}
\usepackage[colorlinks=true, linkcolor=blue, urlcolor=black,citecolor=blue]{hyperref}
\usepackage{orcidlink}
\usepackage{mathtools}

\usepackage{physics}

\newcommand{\one}[0]{\mathds{1}}

\newcommand{\C}{\mathds{C}}

\usepackage{booktabs}
\DeclareUnicodeCharacter{202C}{\^{i}}

\makeatletter
\newcommand{\vast}{\bBigg@{4}}
\newcommand{\Vast}{\bBigg@{5}}
\makeatother
\newtheorem{definition}{Definition}
\newtheorem{theorem}{Theorem}
\newtheorem{conjecture}{Conjecture}

\newtheorem{proposition}{Proposition}
\newtheorem{observation}{Observation}

\usepackage{phfcc}
\phfMakeCommentingCommand[initials={NT}]{nt}

\usepackage{soul,xcolor}

\begin{document}

\title{
	Entangling a quantum system to disentangle its parts
}

\author{Albert Rico$^1$$^{\orcidlink{0000-0001-8211-499X}}$, Jofre Abellanet-Vidal$^2$$^{\orcidlink{0009-0007-3118-0883}}$, Naga Bhavya Teja Kothakonda$^2$$^{\orcidlink{0000-0003-3644-2655}}$, Anna Sanpera$^{2,3}$$^{\orcidlink{0000-0002-8970-6127}}$, Gerard Anglès Munné$^4$$^{\orcidlink{0000-0002-6168-9708}}$}
\affiliation{$^1$Naturwissenschaftlich-Technische Fakult\"{a}t, Universit\"{a}t Siegen, Walter-Flex-Stra\ss e 3, 57068 Siegen, Germany,}
\affiliation{$^2$F\'isica Te\`orica: Informaci\'o i Fen\`omens Qu\`antics, Departament de F\'isica, Universitat Aut\`onoma de Barcelona, E-08193 Bellaterra, Spain,}
\affiliation{$^3$ICREA. Lluis Companys 23, 08010 Barcelona, Spain,}
\affiliation{$^4$Institute of Theoretical Physics and Astrophysics, University of Gdansk, ul.~Wita Stwosza 57, 80-308 Gdansk, Poland.}

\date{\today}
\begin{abstract}
	In quantum systems, correlations are subject to monogamy relations between the whole and its parts. In particular, entanglement between two sets of particles constrains the entanglement within each set and viceversa.
	This raises the following question:
	In a multipartite quantum system, can global correlations disentangle all subsystems of certain size? By exploiting unitary symmetries, we formulate this question as a spectral marginal problem, namely the existence of  $n$-qubit pure states whose marginals below a threshold cannot be entangled by any unitary evolution. We call such states threshold entanglement sharing (TES) states. We provide constructive existence proofs and bound the marginal sizes for up to 200 qubits. 
    Further, we show that
    TES states contain both entanglement and magic, the main resources for quantum computing.
	Our proof strategy bridges techniques that were treated separately up to date, and independently resolves two open problems in quantum information theory: maximal multipartite entanglement and maximum purity of absolutely separable states.
\end{abstract}

\maketitle

Quantum entanglement is the property by which a global physical system does not admit a quantum description based on its parts alone~\cite{epr1935, Bell1964}. 
Sharing entanglement in bipartite systems enables multiple applications, such as computational algorithms displaying quantum advantage~\cite{shor1999factorization,grover1996fastSearch} or secure communication~\cite{Eckert1991EBqkd,Acin2007DIqkd}. Entanglement of multipartite systems has a richer structure~\cite{Dur2000threeQubits,dur_entanglement_2007} and further associated applications ranging from many-body physics to quantum networks, such as distributed quantum computing~\cite{Caleffi_2024_distQCsurvey}, high-precision sensing~\cite{toth2014metrology} or connecting multiple quantum devices~\cite{cacciapuoti2019quantumInternetDistQcomp}.

Entanglement of multipartite quantum systems cannot be shared freely: if two parties are maximally entangled, then none of them can share any 
entanglement with a third party. This phenomenon is known as {\em monogamy of entanglement}~\cite{Terhal_2004monogamy}, and restricts how quantum correlations can be distributed across multipartite systems.
Entanglement monogamy~\cite{Koashi2004Monogamy,Coffman2000DistributEntMonogamy} is responsible for multiple applications, such as security against external adversaries in quantum key distribution~\cite{Eckert1991EBqkd,Acin2007DIqkd}.
Monogamy constraints across bipartitions appear due to the so-called {\em quantum marginal problem}, namely whether given states describing parts of a system can simultaneously be marginals of a global quantum state of the whole system~\cite{klyachko2004quantummarginalproblem,Yu_2021HierMargProb,Klyachko_2006_spectralMarginal,Huber2025SpectralMargProb}.

This leads us to ask how the marginal problem can be formulated if only entanglement correlations are taken into account. That is, how entanglement resources can be distributed between a quantum system and its parts. To understand what entanglement sharing schemes are possible, we ask the following question: {\em Can a quantum system be sufficiently entangled, so that all its parts below a certain threshold size are forced to be disentangled?} 
We denote quantum states enabling this property as threshold entanglement sharing (TES) states.

The purpose behind this question is to understand both the nature of multipartite entanglement and its applications: classical threshold access schemes are methods to share classical information in such a way that no subset of users below a certain threshold size can make use of it. Up to date, the canonical quantum analogue is {\em quantum secret sharing} (QSS)~\cite{Hillery1999QSS}: departing from classical strategies to share a secret~\cite{shamir1979SecretSharing}, entanglement can be used to spread quantum information ({\em qubits}) among multiple users, so that any sufficiently small subset $S$ cannot recover the secret. This is enabled by sharing absolutely maximally entangled (AME) states, whose marginals are maximally mixed and thus contain no information. However, these induce a major obstacle to practical use: their strong constraints impose a rigid structure that is in fact forbidden by quantum mechanics in many system sizes of interest~\cite{Higuchi_2000_twoCouples,Huber2019AME72notExist}. Therefore, one of the applications of solving the question above is to enable a much more flexible framework, where the resource that is unavailable to certain subsystems is quantum entanglement itself (namely {\em e-bits} of information).

\begin{figure}[tbp]
	\centering
	\includegraphics[width=0.9\linewidth]{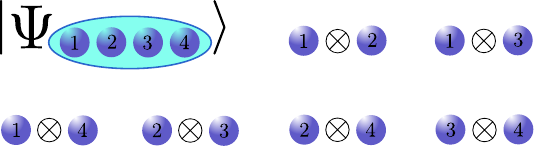}
	\caption{{\bf Threshold entanglement sharing:} we investigate whether a multipartite quantum system can be sufficiently entangled, so that all its marginals below majority are forced to be in a separable quantum state. Monogamy constraints imply that any state with this property must have highly mixed reduced states that remain separable under arbitrary quantum circuit gates. We show that this requirement is remarkably restrictive and, for certain system sizes, cannot be satisfied within quantum mechanics. In the simplest case of $4$ qubits, this can be done with the state of Eq.~\eqref{eq:M4}, where all $2$ qubit marginals are shown to be absolutely separable.}
	\label{fig:TEconcept}
\end{figure}

\bigskip

{\bf \em Pure states with absolutely separable marginals.\,\,}
A bipartite state $\varrho_{AB}$ is separable if it can be described locally as $\varrho_{AB}=\sum_i p_i\varrho_A^i\otimes\varrho_B^i$ with $p_i\geq 0$ and $\sum_i p_i=1$. If this property holds for any global unitary $U\varrho_{AB}U^\dag$, then $\varrho_{AB}$ is called absolutely separable (AS)~\cite{Kus2001ASEP,arunachalam_is_2015,hildebrand_ppt_2007,johnston_separability_2013}. Notice that AS, as it is a unitary invariant property of the state, can only depend on the spectra of the state in hand.
To avoid potential confusion, we remark that the term `absolute' in the notions of AME and AS states is purely coincidental, and both problems have been studied independently up to date~\cite{Rachel2025AMEreview,abellanet-vidal_sufficient_2025}. Multipartite separability is more involved than in the bipartite case, since $r$-partite systems can be separable in different ways. For instance,
the tripartite state $\varrho_A\otimes\varrho_B\otimes\varrho_C$ is fully separable and the state $\varrho_{AB}\otimes\varrho_C$ is separable across the bipartition $AB|C$~\cite{dur_classification_2000, guhne_entanglement_2009}. These differences apply in particular to the notion of absolute separability, whose structure is still far from being fully characterized even in the bipartite scenario~\cite{arunachalam_is_2015,abellanet-vidal_sufficient_2025,johnston_separability_2013}. Here we will denote the set of $r$-qubit states that are AS in all bipartitions between $k$ and $r-k$ qubits as $\text{AS}_{k|r-k}$. For pure $n$-partite states with $n < 8$, considering the bipartition of $k=1$ vs the rest is enough, as we detail in Appendix~\ref{Appendix:separability_structure}.
Absolutely separable states are necessarily mixed states, since all pure (or nearly pure) separable states can be entangled by specific global unitaries. Following this intuition, recent effort has been made to find upper bounds on their purity $\tr(\varrho^2)$~\cite{dung_maximum_2025,ahiable_geometry_2026}.
Let us now take a step back and consider AS states as marginals of $n$-partite pure states $\ket{\psi}$ in a subsystem $S$, i.e.,  $\varrho_S=\tr_{S^c}(\dyad{\psi})$
where the subsystem $S$ and its complementary $S^c$ satisfy $|S|\leq|S^c|$ with $|S|+|S^c|=n$.
Then, the purity of $\varrho_S$ defines in fact a measure of entanglement of the pure state $\ket{\psi}$, known as linear entropy, i.e., $E(\ket{\psi})=1-\tr(\varrho_S^2)$. This hints that AS might admit an interpretation in terms of entanglement monogamy: if a pure state $\ket{\psi}$ is sufficiently entangled across $S|S^c$ to surpass a certain amount of linear entropy of entanglement, then its marginals $\varrho_S$ are sufficiently mixed to be AS. Since the set AS is convex, this amount must be the smallest possible purity among all extremal AS states. A lower bound for this quantity is known to be $(D_S-1)^{-1}$, where $D_S$ is the total dimension of the system $S$~\cite{gurvits_largest_2002}.
Conversely, the largest possible purity of AS states imposes a lower bound on the entanglement of their purification.  This intuition extends to other entanglement measures of pure states.

Our work focuses on understanding this relation between entanglement of the whole and separability of its parts, i.e., pure states and their marginals. This contrasts with earlier approaches to sharing entanglement~\cite{helwig2013AMEexApps,khanian2026entanglementsharingschemes}, which mainly focused on distributing Bell pairs. In particular, we consider the case where entanglement of a multipartite system is shared across all possible bipartitions simultaneously: TES states are $n$-qubit quantum states $\ket{\psi}$, where all bipartitions between a subsystem $S$ and its complement $S^c$ are sufficiently entangled, such that any subsystem $S$ of size $|S|=\lfloor n/2\rfloor$ parties holds a separable state $\varrho_S$. Since the entanglement across $S|S^c$ is invariant under local unitaries of the form $U_S\otimes V_{S^c}$, separability must hold for any rotation $U_S\varrho_S U_S^\dag$. 
Namely, TES states exist if and only if 
there exists a pure state $\ket{\psi}$ with AS marginals $\varrho_S$. This equivalence allows us to define the problem as follows:
\begin{definition}\label{def:TE}
	A threshold entanglement sharing (TES) state is an $n$-qubit pure state $\ket{\psi}$ where all $\lfloor n/2\rfloor$-body marginals are absolutely biseparable. 
\end{definition}
That is, for any subsystem $S$ of size $r=|S|=\lfloor n/2\rfloor$, all marginals $\varrho_S\in\text{AS}_{k|r-k}$ for some $k=1,\dots,\lfloor r/2 \rfloor$.
Notice that $\text{AS}_{k|r-k}$ states are as well $\text{AS}_{1|r-1}$~\cite{hildebrand_ppt_2007, johnston_separability_2013} (see Appendix~\ref{Appendix:separability_structure} for more details).
To better allocate the problem under study, we remark that the definition of TES states can be considered in further generality: one can define $(n,\kappa)$-TES states for a partition of integers $\kappa=[k_1|...|k_\ell]$ with $r:=\sum_{j=1}^\ell k_j$, namely $n$-partite pure states where all of the $r$-body marginals is AS with respect to the partition $\kappa$ of its $r$ local parties. In this broader notion, Definition~\ref{def:TE} reduces to $(n,\kappa=[k|\lfloor n/2\rfloor-k])$-TES states for some $k\leq\lfloor n/2\rfloor$. We will mainly focus on
$(n,\kappa=[1|\lfloor n/2\rfloor-1])$-TES states,
which satisfy that subsystems have access to entanglement between each party and the rest only if their size surpasses the majority. Proving that for certain system sizes these states do not exist will therefore prove that TES states, as defined in Definition~\ref{def:TE}, do not exist either. We focus exclusively on qubit systems since AS conditions for multipartite qudit systems are scarce \cite{hildebrand_entangled_2007, gurvits_separable_2003, gurvits_better_2005, abellanet-vidal_sufficient_2025}.
\bigskip

{\bf {\em Threshold entanglement sharing without secret sharing.\,\,}} 
Remarkably, special instances of TES states are AME states~\cite{helwig2013AMEexApps,Rachel2025AMEreview}, namely pure states where all marginals of size $|S|\leq\lfloor n/2\rfloor$ are maximally mixed, $\varrho_S=\one/2^{|S|}$. Indeed, $U\one U^\dag=\one$ and therefore these remain separable under unitaries. AME states satisfy that all reductions to half systems or less have no information about the global states, and thus allow for so-called {\em quantum secret sharing (QSS)} schemes~\cite{Hillery1999QSS,Helwig2012ameQSS}. However, $n$-qubit AME states exist only for $n=5$ and $n=6$ (besides the trivial cases $n=2$ and $n=3$)~\cite{Huber2019AME72notExist,Higuchi_2000_twoCouples,Scott2004QECentPower}.
For a larger number of qubits, it might be the case that the marginal mixedness imposed by AS is not compatible with a global purification. A natural question is therefore whether TES states can exist when AME states do not. We answer this question positively, showing that TES states fundamentally open further possibilities than standard QSS schemes based on AME states.
\begin{observation}
	TES states exist for system sizes where AME states do not exist.
\end{observation}
We will analyze the paradigmatic cases of four and seven qubits, where AME states do not exist and neither does QSS in the sense of~\cite{Helwig2012ameQSS}.
For the case of $n=4$ qubits~\cite{Higuchi_2000_twoCouples}, consider the following superposition of products of bipartite maximally entangled states,
\begin{align}
	\ket{\phi} & = \frac{1}{\sqrt{6}}\Big (|0011\rangle + |1100\rangle \label{eq:M4} \\
	           & + \omega \left(|0101\rangle + |1010\rangle\right)
	+ \omega^2 \left(|0110\rangle + |1001\rangle\right)\nonumber\Big ),
\end{align}
with $\omega = e^{\frac{2\pi i}{3}}$. This state has appeared earlier in the literature: it is in a sense as entangled as it can be with respect to its bipartitions~\cite{Higuchi_2000_twoCouples} and the geometric measure of entanglement~\cite{Steinberg2024FindingMaxQRes,denker2025chiral}, and it displays quantum advantage at discriminating the parity of permutations~\cite{Diebra2025Parity}. We verify that the marginals of this state
$\varrho_{AB},\,\varrho_{AC},\,\varrho_{AD}$ have the same eigenvalues $\lambda_1=1/2$ and $\lambda_2=\lambda_3=\lambda_4=1/6$, which define two-qubit AS states. Therefore, this state is an example of TES.

Considering each of the four local qubits as a network user, this example contrasts with the non-existence of AME states in $(\C^2)^{\otimes 4}$: for this system size, with Eq.~\eqref{eq:M4} it is possible to prevent any pair of users from sharing e-bits of entanglement by applying circuit gates; while it is well known that it is not possible to prevent that pairs of users share qubits of information by applying circuit gates~\cite{Higuchi_2000_twoCouples}.

This conceptual distinction motivates us to ask whether threshold entanglement sharing is also possible for $n=7$ qubits, where AME states do not exist either~\cite{Huber2019AME72notExist} and hence QSS is not possible~\cite{Helwig2012ameQSS}.
To search examples for this case, we will use that a mixed state (in our case, each marginal) $\varrho_S$ with $|S|=\lfloor n/2\rfloor$ is AS$_{1|\lfloor n/2\rfloor-1}$ if and only if~\cite{hildebrand_ppt_2007, johnston_separability_2013}
\begin{equation}\label{eq:AShildebrand}
	\theta(\vec{\lambda}^S):= \lambda_1^S - \lambda^S_{D-1}-2\sqrt{\lambda^S_{D-2}\lambda^S_D}\leq 0,
\end{equation}
where $D=2^{\lfloor n/2\rfloor}$ and $\lambda^S_i$ are the $i$-th eigenvalues of $\varrho_S$ sorted in non-increasing order. We then consider a gradient descent algorithm starting from random states distributed over the Haar measure~\cite{Zyczkowski2001haar} that, at each iteration, minimizes the cost function
\begin{equation}\label{eq:CostFunctionAS}
	\Theta(\ket{\psi}):=\sum_{|S|=\lfloor n/2\rfloor}\max\big \{0,\theta(\vec{\lambda}^S)\big \}^2.
\end{equation}
Using this algorithm, we find numerical instances of seven-qubit TES states, as well as four-qubit states that are locally inequivalent to Eq.~\eqref{eq:M4}. The code and numerical results are given in~\cite{Code-data}, and relevant properties of a TES example are given in Appendix~\ref{app:propsTE7}.

Alternatively, we also implemented two other methods to find TES states. 
First, we used the algorithm proposed in~\cite{Steinberg2024FindingMaxQRes} and analyzed highly symmetric complex superpositions of the local-Clifford inequivalent graph states proposed in~\cite{Hein2004MultipartyGraph}. 
Second, we adapted a technique that some of the authors used to find AME states~\cite{rico2020absolutely}: given an initial Haar-random state $\ket{\psi}$, we compute its reduction to the first $\lfloor n/2\rfloor$ qubits $\varrho_{1,...,\lfloor n/2\rfloor}$, to obtain $\ket{\psi'}=\varrho_{1,...,\lfloor n/2\rfloor}^{-1/2}\otimes\one\ket{\psi}$, whose reduction to the first $\lfloor n/2\rfloor$ 
qubits is maximally mixed. We normalize and proceed for all possible subsets of $\lfloor n/2\rfloor$ qubits, and repeat iteratively until a state that gives $0$ in the cost function defined in Eq.~\eqref{eq:CostFunctionAS}. 

This algorithm is based on~\cite{Verstraete2003NormalForms}, which is better understood yet serves a different purpose.
We remark that for both methods, neither method yielded any $7$-qubit TES state

\bigskip

{\bf {\em Proofs for non-existence: upper and lower entanglement bounds.\,\,}}
The so-called {\em quantum marginal problem}~\cite{klyachko2004quantummarginalproblem} asks whether there can exist a global quantum state $\varrho$ with specified marginals $\{\varrho_S\}_S$. This problem is relevant for multiple applications in quantum information theory, including the problem of existence of AME states discussed above. Since AS concerns only the eigenvalues of mixed states, a simplified version called {\em spectral marginal problem}~\cite{Klyachko_2006_spectralMarginal,Huber2025SpectralMargProb,Yu_2021HierMargProb} applies to our case: whether there exist pure states $\ket{\psi}$ whose marginals $\varrho_S$ have spectral vector (i.e. vector of eigenvalues) $\vec{\lambda}_S$ compatible with AS conditions. In case such purifications do exist, we are moreover interested in finding upper and lower bounds for the entanglement that they must have.
Namely, assuming an $n$-qubit TES state $\ket{\psi}$ exists, we aim to find what values are possible for its $\lfloor n/2\rfloor$-marginal average purity,
\begin{equation}
	p(\ket{\psi}):= \binom{n}{\lfloor n/2\rfloor}^{-1}\sum_{S:|S|={\lfloor n/2\rfloor}} \tr\big ((\tr_{S^c}\dyad{\psi})^2 \big )\,.
\end{equation}
Lower bounds are given by minimizing the average purity over all possible $n$-partite pure states,
\begin{equation}\label{eq:LBpurity}
	p_{\text{LB}}:=\min_{\ket{\psi}\in{{(\C^2)}}^{\otimes n}} p(\ket{\psi})\,.
\end{equation}
Upper bounds are given by the maximal purity that $|S|=\lfloor n/2\rfloor$-qubit marginal AS states can have,
\begin{equation}\label{eq:UBpurity}
	p_{\text{UB}}:=\max_{\varrho_S\in\text{AS}}\tr(\varrho_S^2)\,
\end{equation}
for a given choice of partition.

By definition, the $\lfloor n/2\rfloor$-marginal average purity $p$ of TES states must be contained in the range $p_{\text{LB}}\leq p \leq p_{\text{UB}}$. In particular, this implies that,
if for a system of $n$ qubits there is a gap, $p_{\text{LB}}>p_{\text{UB}}$, then TES states cannot exist for such number of qubits by contradiction. In fact, this contradiction occurs for $8$ qubits, which means that threshold entanglement sharing is not possible for this system size  (see Table~\ref{tab:Purities}).

The optimizations above have the following operational interpretations: on the one hand, Eq.~\eqref{eq:LBpurity} corresponds to finding the largest possible linear entropy of entanglement, $E(\ket{\psi})=1-\tr\big ((\tr_{S^c}\dyad{\psi})^2\big )$, averaged over all subsystems of size $\lfloor n/2\rfloor$, i.e., $E_{\max}=1-p_{LB}$. It is in fact an open question how high this value can be for a fixed number of qubits~\cite{Facchi2008MaxMultEnt,Facchi_2009StatisticalEnt,Facchi_2010MultEntFrustration}, and multiple entanglement measures in this spirit have been considered~\cite{Scott2004QECentPower,Beckey2021ConcEntMeas,Liu2025GenConcEnt}. AME states upper bound this quantity, since they satisfy $\tr(\varrho_S^2)=2^{-\lfloor n/2\rfloor}$ for all reductions of size $|S|=\lfloor n/2\rfloor$ (if they exist), and therefore the minimization~\eqref{eq:LBpurity} can be seen as the tightest possible approximation to the AME property~\cite{guo2025ApproxKuni}. On the other hand, Eq.~\eqref{eq:UBpurity} gives the maximum Hilbert-Schmidt distance of AS states from the maximally mixed state, $\Delta_{\text{HS}}(\varrho,\one/2^{\lfloor n/2\rfloor})=\sqrt{\tr(\varrho^2)-1/2^{\lfloor n/2\rfloor}}$. From the geometric point of view, it is the radius of the smallest sphere enclosing the set of AS states, and increasing effort has been made to find this outer approximation over the last decades~\cite{hildebrand_ppt_2007,gurvits_better_2005,dung_maximum_2025, xiong2026CritSpec, kondra_fundamental_2026, ahiable_geometry_2026}. Our purpose of bounding the existence of TES states requires solving or finding tight approximations  to the two open problems above, namely the maximal possible entanglement of pure states and the maximal purity of AS states. We do so in the two approaches described below.

\begin{table}[tbp]
	\centering
	\begin{tabular}{c c c c c c c c c c }
		\hline
		$n$                   & 4        & 5     & 6            & 7            & 8
		\\
		\hline
		$p_{\text{UB}}$ & 0.375    & 0.375 & $0.1\Bar{6}$ & $0.1\Bar{6}$ & $0.08\Bar{3}$ 
		\\
		$\hat{p}_{\text{LB}}$
		                      & 0.333(3) & 0.25  & 0.125        & 0.127(8)*    & 0.0857(1)     
		\\
		Exists                & TES      & AME   & AME          & TES          & \textbf{None} 
		\\
		\hline
	\end{tabular}
	\caption{
		{\bf Upper and lower bounds on the $\lfloor n/2\rfloor$-marginal average purity of TES states,}
		resulting from the analytical bound $p_{\rm UB}$ and the rational linear programmig bound $\hat{p}_{\text{LB}}\leq p_{\text{LB}}$, except for $n=7$ for which the result is numerical. For this case, the bound $\hat{p}_{\text{LB}}$ is computed by strengthening the linear programming via SDP (denoted $^*$). The last row summarizes whether TES states exist: if $\hat{p}_{\text{LB}}>p_{\text{UB}}$ for $n$ qubits, then a TES state cannot exist for such system size. This is the case of $8$-qubit TES states. AME states are the only stabilizer instances of TES states, and exist for $5$ and $6$ qubits. Using the numerical gradient descent approach it is possible to find non-AME TES states for $n=5,6$.
        Decimals in brackets are numerical, rounded upwards for $\hat{p}_{\text{LB}}$.
		The analytical bound $p_{\rm UB}$ solves the conjecture posed in~\cite{dung_maximum_2025,ahiable_geometry_2026} for qubits.
	}
	\label{tab:Purities}
\end{table}

\bigskip

\begin{figure}
	\centering
	\includegraphics[width=\linewidth]{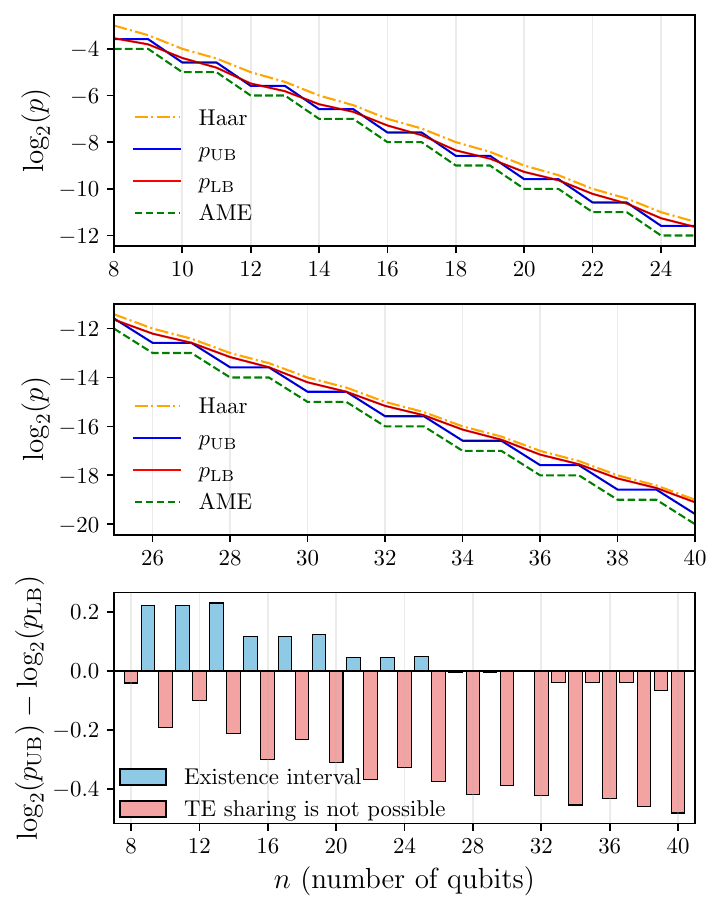}
	\caption{{\bf Existence of TES states between $8$ and $40$ qubits.} {\em Top:} We compare two independent results found in this work: the lower bounds found numerically in this work with the LP from Eq.~\eqref{eq:lp} (Approach 1), and the upper bounds $p_{\text{UB}}$ given by the maximal purity of absolutely separable states (Approach 2). The logarithm is taken for better visibility.
		For the regime analyzed, TES states are more entangled than random states distributed over the Haar measure.
		The comparison to AME states is made for completeness, although these do not exist for $7$ qubits~\cite{Huber2019AME72notExist} and beyond~\cite{Scott2004QECentPower}. {\em Bottom:} We compute the difference between logarithms of the upper ($p_{\text{UB}}$) and lower ($p_{\text{LB}}$) bounds of the marginal average purity $p$.
		According to the data, TES states could exist for $n$ odd up to $n=25$, but did not for $n$ even. Even though we only show results up to $n=40$, we compute the LP until $n=200$ and demonstrating that TES states do not exist for $26 \leq n\leq 200$.
		We recall that all computations are exact since Approach 1 gives an optimal rational solution and Approach 2 ensures an analytical bound.
	}
	\label{fig:Puritycombined}
\end{figure}

{\bf \em Approach 1. Minimal purity of marginals.\,\,} 
We first aim to lower-bound Eq.~\eqref{eq:LBpurity}.
To this end, we will view TES states as one-dimensional almost quantum codes and adapt the optimization machinery of quantum coding theory to our problem~\cite{PhysRevLett.78.1600,681316}. For that, we briefly recall the following property of \emph{pure quantum codes} $C$ capable of correcting $e$ errors: these encode a $K$-dimensional Hilbert space to a (code) subspace of $n$-qubit Hilbert space, such that all errors $E$ acting nontrivially on $e$ or fewer qubits and all states $\ket{\psi}$ from that subspace satisfy $\bra{\psi}E\ket{\psi}=0$. This requires that all $m$-body marginals of $\ket{\psi}$ are maximally mixed, namely $\varrho_S=\one/2^m$ if $|S|=m$. Pure states whose marginals have this property are known as {\em $m$-uniform states}~\cite{Goyeneche2015AMEcombDes}. Thus, all $\ket{\psi}$ are maximally entangled between any $m$ qubits and the rest.
For our purposes, we do not need this property; instead, we need a high degree of mixedness on the marginals. That is, we have $\bra{\psi}E\ket{\psi}\approx 0$, which allows us to regard TES states $\ket{\psi}$ as elements of almost pure quantum codes.
Since it is one-dimensional, the subspace contains only one approximate $m$-uniform state $\ket{\psi}$~\cite{guo2025ApproxKuni}. 
Therefore, we will apply to our problem the following optimization variant of the feasibility linear programs (LP) used in~\cite{PhysRevLett.78.1600,681315,681316,796376} to 
determine the existence of one-dimensional quantum codes 
\begin{theorem}\label{th:lb}
	The following bound holds, $\hat{p}_{\rm LB}\leq p_{\rm LB}$, where $\hat{p}_{\rm LB}$ can be computed with linear programming as follows: 
	\begin{align}
		\hat{p}_{{\rm LB}} := \min_{\{a'_j\}_{j=0}^{n}} \quad & \binom{n}{\lfloor n/2\rfloor}^{-1}  a'_{\lfloor n/2\rfloor}   \label{eq:lp}    \\
		{\rm s.\,t. }	\quad                                   & a'_0=1,\quad a'_j=a'_{n-j}\,,\nonumber                                         \\
		                                                      & s_j :=\sum^n_{\ell=0} K_{n-j}(\ell) a'_\ell \,\geq 0 \,\nonumber                       \\
		                                                      & a_j:=\sum^j_{\ell=0} (-1)^{j-\ell}2^{\ell} \binom{n-\ell}{n-j}a'_\ell \,\geq \,0 \nonumber\, ,
	\end{align}

	for all subset sizes $j=0,\dots, n$ and $s_j=0$ for $n-j$ odd, where $K_u(x)=\sum^{n}_{\alpha=0} \binom{n-x}{u-\alpha}\binom{x}{\alpha}(-1)^\alpha$ are the \emph{Krawtchouk polynomials}.
\end{theorem}
Here the variables $a'_j$ are a relaxation of the sum of purities of the marginals of $\ket{\psi}$ for a given set size $j$,
\begin{align}
	A'_j= \sum_{S:|S|= j} \tr\big ((\tr_{S^c}(\dyad{\psi})^2 \big ) ,
\end{align}
where the notation $A_j'$ follows standard quantum error correction notation for code projectors~\cite{681316}. The proof and further details about the code machinery used are explained in Appendix~\ref{app:minpurity}. Theorem~\ref{th:lb} allows to systematically compute $p_{\text{LB}}$ for large system sizes (Figure~\ref{fig:Puritycombined}-Top), which is currently an open problem~\cite{Facchi_2010MultEntFrustration,Facchi2008MaxMultEnt}. Indeed, to our knowledge, this provides the tightest existing bound for the maximal average entanglement of pure states, $E_{\max}=1-p_{\text{LB}}$~\cite{guo2025ApproxKuni,Beckey2021ConcEntMeas,Facchi_2009StatisticalEnt}.

Notice that the smallest possible value of $\hat{p}_{\text{LB}}$ is $2^{-\lfloor n/2\rfloor}$, which must be attained if an $n$-qubit AME state exists. The results of this optimization are given in Table~\ref{tab:Purities}.
Tighter bounds than those given by the LP of Theorem~\ref{th:lb} can be obtained by employing the SDP formulation on self-dual quantum codes derived in~\cite{munne2025sdpboundsquantumcodes,munne2026sdpboundsquantumcodes}.
Departing from the LP, we adapt the SDP constraints for quantum codes introduced in~\cite[Eq.~(31)]{munne2026sdpboundsquantumcodes}, which were designed for AME states, to the more general case of TES states; and maximize over $a'_{\lfloor n/2\rfloor}$ considering that the sum of purities $A'_j$ can also be expressed in terms of the $A_i$ enumerators~\cite[Corollary 5]{681316}.
This SDP formulation is strictly stronger than the LP formulation above. Indeed, for $n=7$ qubits, the LP gives $\hat{p}_{\text{LB}}=0.125$, i.e., it cannot disprove the existence of the seven-qubit AME states, while the SDP formulation does~\cite{munne2025sdpboundsquantumcodes}.
However, SDPs are generally harder to compute compared to LPs.
The SDP computations were carried out using Clarabel~\cite{Clarabel_2024} through JuMP~\cite{jump}, with primal and dual feasibility tolerances set to $10^{-10}$.
The resulting SDP bound in average purity is denoted as $^*$ in Table~\ref{tab:Purities}.
Differently, for the LP, we use SageMath~\cite{sagemath} with its Parma Polyhedra Library~\cite{ppl} (PPL) backend.
This allows us to find an exact solution of the LP, i.e., the primal and dual solutions are rational and satisfy strong duality~\cite{boyd}.
Thus, the values $\hat p_{LB}$ from the LP are rational and certified to be optimal, avoiding numerical imprecision.

Using the mentioned solver, we could compute $p_{\text{LB}}$ up to $n=200$ qubits. In Figure~\ref{fig:Puritycombined}, we plot our results for $8\leq n\leq 40$, which is enough to observe that $\log_2(p_{\text{LB}})$ decreases nearly linearly with the even and odd number of qubits $n$. The results of the LP and the code used are given in~\cite{Code-data}.

Although this analysis deviates from the main purpose of this work, it is to our knowledge the tightest known estimate of the maximal pure-state entanglement averaged over bipartitions for multiqubit systems~\cite{Beckey2021ConcEntMeas,Liu2025GenConcEnt,guo2025ApproxKuni,Facchi2008MaxMultEnt,Facchi_2009StatisticalEnt,Facchi_2010MultEntFrustration}.

\bigskip

{\bf \em Approach 2. Maximal purity of AS states.\,\,}
The second approach we follow, in combination with the previous one, is to upper bound Eq.~\eqref{eq:UBpurity}. The maximal purity of the set of AS states has been extensively studied \cite{gurvits_largest_2002, filippov_absolutely_2017, dung_maximum_2025, Song_2025_Extreme_APPT,xiong2026CritSpec,kondra_fundamental_2026, ahiable_geometry_2026}, but no tight upper bound has been proven. Here, we have been able to prove Conjecture 3.1 of \cite{dung_maximum_2025}, and Conjecture 6.7 of \cite{ahiable_geometry_2026} (for the qubit-qudit case) for our case of interest of $r$-qubit systems, considering absolute separability in the partition of $1$ vs the rest.
\begin{theorem}\label{th:ub}
	The maximal purity of an absolutely separable state $\varrho$ in $\mathbb{C}^{2}\otimes\mathbb{C}^{2^{r-1}}$ for $r\geq3$ is
	\begin{equation}
		\label{eq:maximal_purity}
		\max_{\varrho\in AS}\tr(\varrho^{2}) = \frac{4}{3\cdot 2^{r}}.
	\end{equation}
\end{theorem}
The detailed proof for $r$-qubit systems is provided in Appendix~\ref{App:ProofMaxPurityAS2d}. Besides independently solving an existing open problem, this allows us to address the non-existence of TES states. 
Up to date, this was an open conjecture and had only been proven for the case of $r=2$~\cite{dung_maximum_2025}.

\bigskip

{\bf {\em Non-existence of threshold entanglement sharing.\,\,}}
The results of the previous sections allow us to address the non-existence of TES states.
For our purposes, we compare the rational bounds we found for up to $n=200$ qubits via Approach 1 and the analytical bounds of Approach 2. Together with the TES states found algorithmically, we collect the results found with the methods above and answer the question posed in this work. Denote $\mathcal{N}_{\exists}:=\{4, 5, 6, 7\}$, and $\mathcal{N}_{\not\exists}:=\{n_{\rm even}\,:\,8\leq n\leq 24\}\cup\{n\,:\,26\leq n\leq 200\}$, where $n_{\rm even}$ is the even number of qubits $n$.
\begin{theorem}
	Threshold entanglement in $n$-qubit systems is possible for $n\in\mathcal{N}_{\exists}$, and impossible for $n\in\mathcal{N}_{\not\exists}$. That is:
	\begin{enumerate}
		\item  For $n\in\mathcal{N}_{\exists}$, there exist pure quantum states $\ket{\psi}\in(\C^2)^{\otimes n}$ where, in every marginal of size $\lfloor n/2\rfloor$, no qubit can be entangled to the rest with unitary evolutions.
		\item For $n\in\mathcal{N}_{\not\exists}$, every pure state $\ket{\psi}\in(\C^2)^{\otimes n}$ has at least one subsystem of size $\lfloor n/2\rfloor$ where one qubit can be entangled to the rest with unitary evolutions.
	\end{enumerate}
\end{theorem}
Notice that, due to the unitary invariance, separability of the marginals is induced solely from the correlations between any subsystem of $\lfloor n/2\rfloor$ qubits and the rest. In this sense, the results for $n\in\mathcal{N}_{\not\exists}$ prove entanglement frustration, namely that quantum mechanics does not allow sufficiently strong multipartite correlations to disentangle the marginals of half size.
Notably, none of the approaches combined could answer whether this is possible for an odd number of qubits $9\leq n_{\rm odd}\leq 25$. We attribute this to the fact that our algorithms to find TES states (and all algorithms to find specific states we are aware of~\cite{Steinberg2024FindingMaxQRes,Rather2022AME46}) are not scalable, in the sense that they require storing $n$-qubit quantum states in a classical computer. However, for this range of $n$, we estimate the intervals of logarithmic purities (equivalently, linear entropy of entanglement) that these states must have in order to exist (blue bars in Figure~\ref{fig:Puritycombined}-Bottom). This interval is compressed inside $p_{\rm UB}-p_{\rm LB}\leq 0.012$ for such system sizes. Non-existence is proven by the contradiction $\hat{p}_{\text{LB}}>p_{\text{UB}}$ (red bars in Figure~\ref{fig:Puritycombined}-Bottom). The pattern observed up to $n=200$ suggest that TES states cannot exist for any number of qubits $n\geq 26$. The non-existence of marginals with more restrictive separability structures is discussed in  Appendix~\ref{Appendix:separability_structure}.

\bigskip

{\bf {\em Marginals below majority.\,\,}}
The above results shed light on the possibility of sharing entanglement that is inaccessible to subsystems below majority. We gained evidence that this is not possible for $n\geq 26$ qubits, but this is due to the choice made of restricting all marginals of size $\lfloor n/2\rfloor$ to be separable up to unitary evolutions.
Here, we relax the definition of threshold entanglement to the following

\begin{definition}\label{def:mTE}
	An $m$-TES state of $n$ qubits is an $(n,[k,m-k])$-TES state. That is, an $n$-qubit pure state whose $m$-body marginals are absolutely separable with respect to the partition $[k,m-k]$.
    \end{definition}

\begin{figure}
	\centering
	\includegraphics[width=\linewidth]{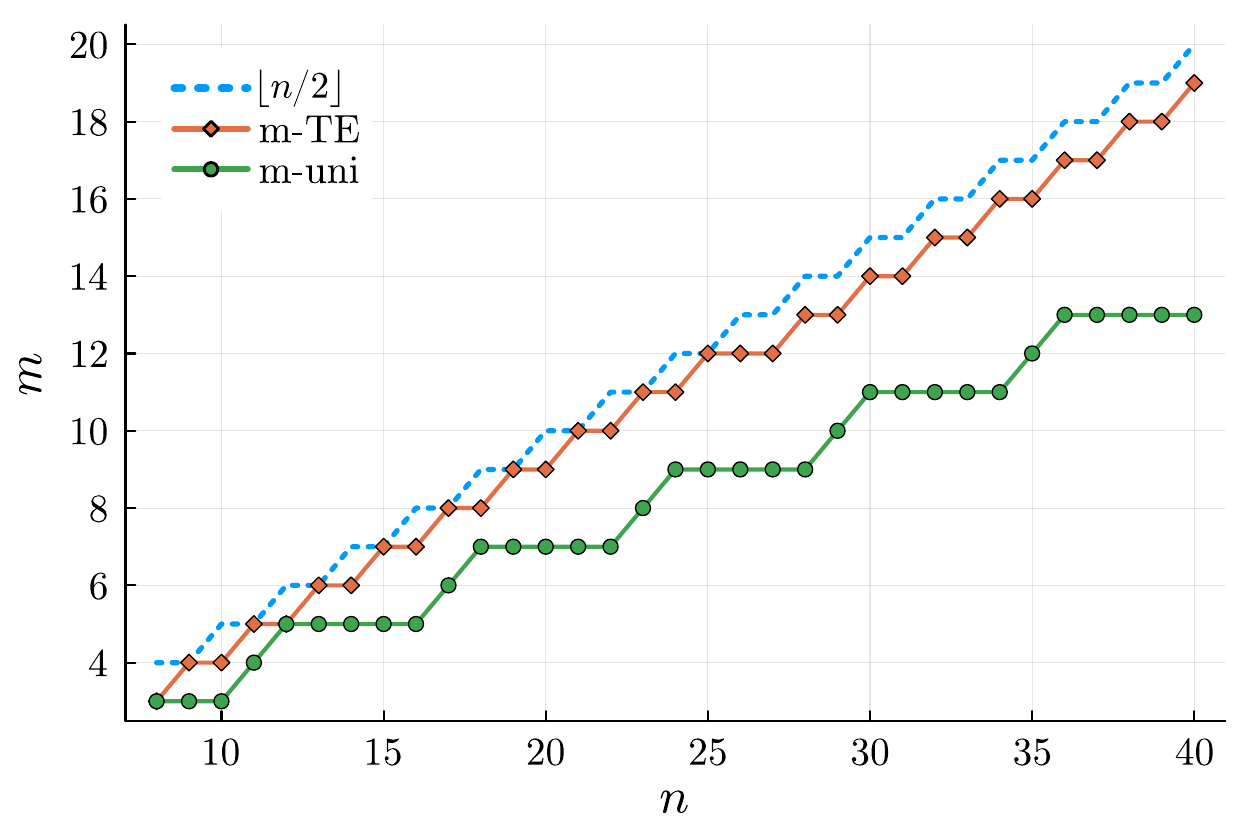}
	\caption{\textbf{Existence of $m$-TES states between $8$ and $40$ qubits.} The plot shows the maximum possible $m$ for a given $n$ such that an $m$-TES state of an $n$-qubit system could exist (orange). The result is upper-bounded by $m=\lfloor n/2 \rfloor$ (dashed) and lower-bounded by the maximum $m$ such that an $m$-uniform state of $n$ qubits could exist \cite{Goyeneche2015AMEcombDes} (green).}
	\label{fig:maxm}
\end{figure}

In the same way that AME states are obvious examples satisfying Definition~\ref{def:TE}, $m$-uniform states are examples of $m$-TES states in Definition~\ref{def:mTE}. In this sense, our next goal is to gain insight into the non-existence of $m$-TES states: we compute the maximum $m$ such that there can potentially exist an $m$-TES state, in the sense that it does not contradict the conditions for AS in Theorem~\ref{th:ub}.
The maximum $m$ is determined when, for a given state, there exists a feasible solution  of Theorem~\ref{th:lb}, $\{a'_j\}^n_{j=0}$, additionally satisfying
\begin{align}\label{eq:ASlp}
	\frac{4}{3 \cdot 2^r} \binom{n}{r} \geq a'_r  \quad \text{for} \quad r\leq m \, .
\end{align} 
but not being satisfied for $r=m+1$.
The results are illustrated in Figure~\ref{fig:maxm} for $8 \leq n \leq 40$.
These are compared with the maximum $m$ for which there exists a pure state in which all $m$-body marginals are maximally mixed (i.e. $m$-uniform states~\cite{Goyeneche2015AMEcombDes}).
We can check the existence of these states by converting the inequalities in Eq.~\eqref{eq:ASlp} to equalities and adding the resulting constraints to the feasibility version of LP~\eqref{eq:lp}.
This gives us an upper bound on the existence of $m$-uniform states.
Even though for  $8\leq n \leq26$, TES states with even $n$ do not exist, we found that the LP solutions of Theorem~\ref{th:lb} are also compatible solutions $\{a'_j\}$ with the existence of $(\lfloor n/2 \rfloor -1)$-TES states.
Surprisingly, from $n=26$ to $n=$200 for both even and odd, the maximum possible $m$ is always $\lfloor n/2 \rfloor -1 $. 
This is because the LP of Theorem~\ref{th:lb} already gives us a rational feasible solution up to $n=200$ additionally satisfying Eq.~\eqref{eq:ASlp} for $m=\lfloor n/2 \rfloor -1 $.
Thus, states might potentially exist for any number of qubits.

\bigskip
{\bf \em Quantum computing resources of TES sharing.\,\,}
Two main resources are needed for quantum computing protocols that cannot be simulated efficiently by classical computers: entanglement and non-stabilizerness~\cite{Briegel2003MBQcomp,gottesman1998HeisClasSimStab}. TES states are by definition highly entangled, and in fact Approach 2 provides lower bounds on their linear entropy of entanglement $E_{\min}:=1-p_{\rm UB}$. A natural question is therefore whether they must also be non-stabilizers. While AME states can be implemented via stabilizer graph states~\cite{helwig2013AmeGraph,Briegel2003MBQcomp,gottesman1997stabilizer}, this is not the case for other system sizes:
\begin{observation}
	For any number of qubits $n$ satisfying $n\neq 5$ and $n\neq 6$, TES states cannot be stabilizer states.
\end{observation}

This is because the partial trace of a stabilizer state is the projector onto a stabilizer code (up to normalization), whose rank is an integer power of two, $2^s$; but the rank of $\lfloor n/2\rfloor$-partite AS mixed states is either $2^{\lfloor n/2\rfloor}$ or $2^{\lfloor n/2\rfloor}-1$~\cite{hildebrand_ppt_2007, johnston_separability_2013, abellanet-vidal_sufficient_2025}, and therefore a TES stabilizer state must either be AME or not exist.

The next question is then how much of non-stabilizerness resource they can accommodate. This resource is known as {\em magic}~\cite{Howard2017Magic} and for qubit systems it can be quantified by the stabilizer $\alpha$-Rényi entropy,
\begin{equation}\label{eq:MagicRenyi}
	S_\alpha(\ket{\psi}) = \frac{1}{1-\alpha}\log_2\Bigg ( \sum_{P\in\mathcal{P}_n} \frac{\bra{\psi}P\ket{\psi}^{2\alpha}}{2^n} \Bigg )
\end{equation}
with $\alpha \neq 1$, $\alpha>0$, and where $\mathcal{P}_n$ is the set of $n$-qubit Pauli strings, e.g. $P=\sigma_X\otimes\sigma_Y\otimes\one\otimes\sigma_Z$, modulo phases $(\pm 1,\pm i)$~\cite{Howard2017Magic,sierant2026computingMagic}. Rényi entropies recover the Shannon entropy in the limit $\alpha\rightarrow 1$.\\

We will particularly focus on the amount of magic (considering $\alpha=2$) present in the examples of TES states we find by using the algorithm minimizing Eq.~\eqref{eq:CostFunctionAS}, in comparison to generic input states. First, let us define the average amount of magic for a given sample $\{\psi_i\}^N_{i=1}$ of size $N$ as $\langle S_2(\ket{\psi})\rangle_{N}:=\tfrac{1}{N}\sum_{i=1}^{N}S_2(\ket{\psi_i})$. This value can be bounded by $0\leq S_2(\ket{\psi})\leq \log_2(2^n+1)-1$ for $n$-qubit pure quantum states~\cite{Leone2022StabREnt}. \\

Starting from $2\cdot 10^3$ Haar-random states, we obtain a sample of $1995$ $4$-qubit TES states $\{\ket{\psi_i^4}\}_{i=1}^{10^3}$ whose average magic is $\langle S_2(\ket{\psi^4})\rangle_{1995}\approx 2.08 \pm 0.18$, where $\pm c$ is the standard deviation of $S_2$ over the set of states.
This is a significant amount compared to the upper bound for $4$-qubit systems, which reads $\log_2(2^4+1)-1\approx 3.08$.
In comparison, we verify that $\langle S_2(\ket{H^4})\rangle_{2\cdot10^3}\approx 2.25\pm0.13$ over $4$-qubit Haar-random states $\ket{H^4}$, which are known to accommodate high amounts of magic as well as entanglement~\cite{Turkeshi2025MagicTypical}.
Using the same algorithm but starting with $2\cdot 10^3$ random stabilizer states instead of Haar-random, we could only generate a sample of $71$ TES states $\ket{\phi^4}$ with $\langle S_2(\ket{\phi^4})\rangle_{{71}}\approx 1.58\pm0.4$.
For a starting state with zero magic, it is more difficult to obtain TES states using the proposed algorithm and the ones obtained have less magic.
In fact, we observe that the analytical state $\ket{\phi}$ in Eq.~\eqref{eq:M4} has a magic value closer to that of TES states obtained from stabilizer states: preliminary inspection shows that it is stabilized by the two Pauli strings $\sigma_X^{\otimes 4}$ and $\sigma_Z^{\otimes 4}$ generating a stabilizer subgroup of $\mathcal{P}_4$, and thus, 
$S_2(\ket{\phi})= \log_2(9/4)\approx 1.17$. \\

\begin{figure}
	\centering
	\includegraphics[width=\linewidth]{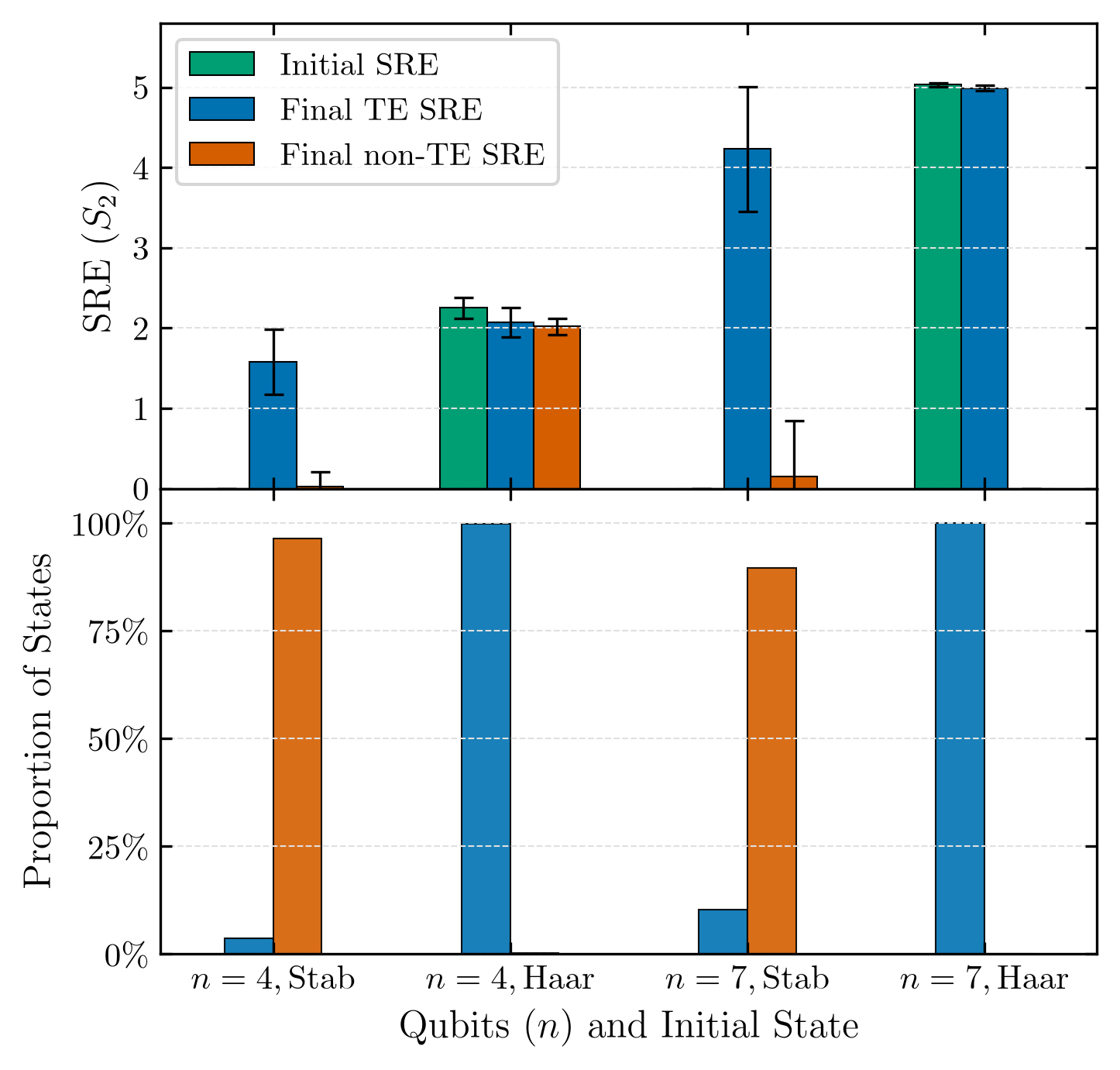}
	\caption{{\bf Comparing $S_2$ of TES with non-TES:} In the above chart we consider two initializations for the optimization algorithm starting with the same number $N=2\cdot 10^3$  random Stabilizer states (Stab) and Haar random states (Haar) for all the system sizes. We run the optimization algorithm for both of the initializations for the same number of steps i.e. till $t=100$ for $4$-qubit systems and $t=1500$ for $7$-qubit systems. For the Stab initialized system, SRE of TES is significantly higher compared to SRE of the non-TES. This difference is much smaller for the Haar initialized systems. See Appendix~\ref{app:Magic} for greater detail. }
	\label{tab:MagicPurity}
\end{figure}

Similarly, for the case of $7$ qubits, we have generated a sample of $2\cdot 10^3$ TES states $\ket{\psi^7}$, by minimizing Eq.~\eqref{eq:CostFunctionAS} with $2\cdot 10^3$ $7$-qubit Haar random states as seeds. The average amount of magic over the TES states found is $\langle S_2(\ket{\psi^7})\rangle_{2\cdot10^3}\approx 4.99\pm0.03$, in comparison to $7$-qubit Haar random states with average magic $\langle S_2(\ket{H^7})\rangle_{2\cdot10^3}\approx 5.03\pm0.03$. The upper bound for $7$-qubit pure states given by $\log_2(2^7+1)-1\approx 6.02$.
In contrast, when initial states are stabilizers, the number of TES states obtained $\ket{\phi^7}$ is $207$ and $\langle S_2(\ket{\phi^7})_{207}\approx 4.23\pm 0.78$.
Note that for the $7$-qubit state, the magic after running the gradient descent algorithm increases significantly from zero.
This fact suggests that TES states require a considerable amount of magic.
All results are summarized in Figure~\ref{tab:MagicPurity}.
\bigskip

These numerical observations suggest that TES states are highly resourceful not only in entanglement, but also in magic. Therefore, we expect that protocols using threshold entanglement sharing might be candidates for finding quantum computing tasks that cannot be efficiently simulated by classical computers.
Further details on the numerics are given in Appendix~\ref{app:Magic}.

\bigskip

{\bf \em Conclusions.\,\,}
Here, we have addressed the question of whether, given a sufficiently entangled multipartite system, some of its bipartite partitions are necessarily forced to be disentangled due to monogamy constraints. Using the fact that such subsystems cannot be entangled by means of unitary transformations (quantum circuit gates or Hamiltonian evolutions), we have formalized this question in terms of {\em threshold-entanglement sharing (TES)} states, namely, multipartite pure states whose marginals below the majority threshold are absolutely separable mixed states. In this respect, TES states can be regarded as generalizations of absolutely maximally entangled (AME) states. We have shown that TES states exist for system sizes for which AME states do not, opening the way to novel entanglement-threshold schemes. 
 For systems of up to $200$ qubits, we have determined the range of average entanglement, $E_{\min}$ and $E_{\max}$, required for TES states. To this end, we have computed lower bounds on the smallest possible marginal purity of pure states and upper bounds on the largest possible purity of absolutely separable states. In doing so, our strategy addresses two important open problems in quantum information theory. First, we have derived systematic upper bounds on the maximal average entanglement achievable in $n$-qubit systems within quantum theory. Second, we have confirmed an open conjecture concerning the boundary of the set of absolutely separable states with respect to the bipartition between one qubit and the rest. By combining these two results, we have shown that qubit TES states cannot exist for certain system sizes. Furthermore, we have demonstrated that TES states possess significant amounts of both entanglement and magic, two essential resources for quantum computational protocols that cannot be efficiently simulated classically. 
 Thus, by exploring fundamental limitations imposed by genuinely multipartite entanglement, we offer a novel perspective on the distribution of quantum information, focusing on e-bits rather than qubits. Remarkably, this opens a new avenue for exploring cryptographic strategies that are considerably less demanding than current schemes, such as quantum secret sharing, which is known to become unfeasible beyond systems of only a few qubits. 
 
 A potential application of TES states relies on the distribution of entanglement across quantum networks, where sets of users below the majority threshold, cannot perform quantum protocols without cooperating with other users. However, the development and analysis of specific protocols demand further investigation. Finally, we emphasize that the existence of TES states beyond qubits remains a challenging open question, since conditions ensuring absolute separability in multipartite qudit systems are largely unknown.

\bigskip

{\em Acknowledgements.}
We are thankful to Jennifer Ahiable, Dagmar Bruß, Felix Huber, Kenneth Goodenough, Otfried Gühne, Andreas Winter, and Karol Życzkowski for comments and discussions. We acknowledge the use of a computational server financed by the Foundation for Polish Science (IRAP project, ICTQT, contract no. 2018/MAB/5/AS-1, co-financed by EU within Smart Growth Operational Programme. AR, JAV, NBTK, AS acknowledge
support from  PID2022-141283NB-I00 with the support of FEDER funds.
AR, JAV, NBTK and AS acknowledge also financial support  from the Spanish
Government with funding from European Union
NextGenerationEU (PRTR-C17.I1), the Generalitat
de Catalunya, the Ministry for Digital
Transformation and of Civil Service of the Spanish
Government through the QUANTUM ENIA
project -Quantum Spain Project- through the
Recovery, Transformation and Resilience Plan
NextGeneration EU.
AR acknowledges support
by the Deutsche Forschungsgemeinschaft, DFG, German
Research Foundation, project number 563437167 and Project BeRyQC, Grant No.
13N17292. JAV acknowledges financial support from Ministerio de Ciencia e Innovación of the Spanish Government FPU23/02761. NBTK acknowledges partial support by the Alexander von Humboldt Foundation.  GAM was supported by NCN grant no. 2024/53/B/ST2/04103.

\newpage
\bigskip

\addcontentsline{toc}{subsection}{Bibliography}
\bibliographystyle{ieeetr}
\bibliography{Bibliography}{}

\newpage
\begin{widetext}
	\appendix
	\section{Constructions and properties of TES states}
	\label{appendix:Example7Qubits}\label{app:propsTE7}
	In this appendix, we introduce a particular example of an unnormalized $7$-qubit TES state $|\tilde \phi \rangle$ obtained with the gradient descent method detailed in Eq. \eqref{eq:CostFunctionAS}.
	Note that the normalization does not play a role in determining if the state is TES since the AS condition of Eq.~\eqref{eq:AShildebrand} is independent of the normalization.
	The numerical result has been rationalized using the SymPy package for Python \cite{noauthor_peerj_nodate}  to obtain an analytical instance.
    
	\begin{equation}
		\resizebox{0.86\textwidth}{!}{%
			$\displaystyle
				|\tilde \phi \rangle=
				\begin{array}[t]{@{}r@{\;}l@{\quad}r@{\;}l@{\quad}r@{\;}l@{\quad}r@{\;}l@{}}
				             {}-\left(\frac{3}{40}+\frac{2}{105}\,\mathrm{i}\right) & \ket{0000000} & {}-\left(\frac{7}{97}+\frac{9}{91}\,\mathrm{i}\right)    & \ket{0000001} & {}+\left(\frac{7}{159}-\frac{2}{45}\,\mathrm{i}\right)    & \ket{0000010} & {}-\left(\frac{7}{127}-\frac{8}{115}\,\mathrm{i}\right)   & \ket{0000011} \\
				             {}+\left(\frac{1}{30}+\frac{5}{88}\,\mathrm{i}\right)           & \ket{0000100} & {}+\left(\frac{3}{121}-\frac{1}{23}\,\mathrm{i}\right)   & \ket{0000101} & {}-\left(\frac{5}{78}-\frac{1}{25}\,\mathrm{i}\right)     & \ket{0000110} & {}+\left(\frac{1}{79}+\frac{7}{72}\,\mathrm{i}\right)     & \ket{0000111}     \\
				             {}+\left(\frac{3}{107}-\frac{2}{39}\,\mathrm{i}\right)          & \ket{0001000} & {}-\left(\frac{7}{64}-\frac{3}{88}\,\mathrm{i}\right)    & \ket{0001001} & {}-\left(\frac{10}{123}-\frac{10}{161}\,\mathrm{i}\right) & \ket{0001010} & {}-\left(\frac{2}{51}+\frac{1}{60}\,\mathrm{i}\right)     & \ket{0001011}     \\
				             {}-\left(\frac{7}{67}+\frac{7}{79}\,\mathrm{i}\right)           & \ket{0001100} & {}+\left(\frac{1}{69}-\frac{4}{55}\,\mathrm{i}\right)    & \ket{0001101} & {}-\left(\frac{14}{109}+\frac{3}{100}\,\mathrm{i}\right)  & \ket{0001110} & {}+\left(\frac{1}{17}+\frac{5}{61}\,\mathrm{i}\right)     & \ket{0001111}     \\
				             {}-\left(\frac{5}{37}+\frac{1}{37}\,\mathrm{i}\right)           & \ket{0010000} & {}-\left(\frac{6}{113}-\frac{1}{14}\,\mathrm{i}\right)   & \ket{0010001} & {}+\left(\frac{1}{11}+\frac{4}{59}\,\mathrm{i}\right)     & \ket{0010010} & {}-\left(\frac{4}{81}-\frac{7}{114}\,\mathrm{i}\right)    & \ket{0010011}     \\
				             {}-\left(\frac{2}{67}+\frac{2}{91}\,\mathrm{i}\right)           & \ket{0010100} & {}-\left(\frac{2}{59}+\frac{14}{169}\,\mathrm{i}\right)  & \ket{0010101} & {}+\left(\frac{1}{104}-\frac{5}{126}\,\mathrm{i}\right)   & \ket{0010110} & {}+\left(\frac{7}{99}-\frac{2}{163}\,\mathrm{i}\right)    & \ket{0010111}     \\
				             {}-\left(\frac{3}{44}-\frac{3}{103}\,\mathrm{i}\right)          & \ket{0011000} & {}+\left(\frac{10}{103}-\frac{5}{93}\,\mathrm{i}\right)  & \ket{0011001} & {}-\left(\frac{1}{89}-\frac{2}{103}\,\mathrm{i}\right)    & \ket{0011010} & {}-\frac{1}{83}                                           & \ket{0011011}     \\
				             {}-\frac{5}{107}\,\mathrm{i}                                    & \ket{0011100} & {}+\left(\frac{8}{97}-\frac{1}{86}\,\mathrm{i}\right)    & \ket{0011101} & {}+\left(\frac{2}{99}+\frac{2}{33}\,\mathrm{i}\right)     & \ket{0011110} & {}+\left(\frac{11}{109}-\frac{5}{101}\,\mathrm{i}\right)  & \ket{0011111}     \\
				             {}-\left(\frac{6}{91}-\frac{6}{149}\,\mathrm{i}\right)          & \ket{0100000} & {}+\left(\frac{4}{59}-\frac{9}{92}\,\mathrm{i}\right)    & \ket{0100001} & {}-\left(\frac{8}{123}-\frac{1}{17}\,\mathrm{i}\right)    & \ket{0100010} & {}-\left(\frac{5}{76}+\frac{1}{121}\,\mathrm{i}\right)    & \ket{0100011}     \\
				             {}-\left(\frac{7}{90}+\frac{1}{170}\,\mathrm{i}\right)          & \ket{0100100} & {}-\left(\frac{2}{113}-\frac{1}{80}\,\mathrm{i}\right)   & \ket{0100101} & {}-\left(\frac{1}{26}+\frac{4}{131}\,\mathrm{i}\right)    & \ket{0100110} & {}-\left(\frac{12}{191}-\frac{13}{132}\,\mathrm{i}\right) & \ket{0100111}     \\
				             {}-\left(\frac{13}{83}+\frac{1}{29}\,\mathrm{i}\right)          & \ket{0101000} & {}-\left(\frac{2}{145}-\frac{6}{67}\,\mathrm{i}\right)   & \ket{0101001} & {}+\frac{3}{113}\,\mathrm{i}                              & \ket{0101010} & {}-\left(\frac{17}{171}+\frac{5}{73}\,\mathrm{i}\right)   & \ket{0101011}     \\
				             {}+\left(\frac{1}{65}+\frac{10}{129}\,\mathrm{i}\right)         & \ket{0101100} & {}+\left(\frac{2}{141}-\frac{1}{128}\,\mathrm{i}\right)  & \ket{0101101} & {}+\left(\frac{5}{132}-\frac{3}{44}\,\mathrm{i}\right)    & \ket{0101110} & {}+\left(\frac{15}{142}-\frac{1}{49}\,\mathrm{i}\right)   & \ket{0101111}     \\
				             {}+\left(\frac{3}{86}-\frac{7}{143}\,\mathrm{i}\right)          & \ket{0110000} & {}-\left(\frac{1}{69}-\frac{1}{50}\,\mathrm{i}\right)    & \ket{0110001} & {}+\left(\frac{3}{55}-\frac{10}{103}\,\mathrm{i}\right)   & \ket{0110010} & {}+\left(\frac{6}{67}+\frac{1}{72}\,\mathrm{i}\right)     & \ket{0110011}     \\
				             {}-\left(\frac{7}{109}-\frac{11}{89}\,\mathrm{i}\right)         & \ket{0110100} & {}+\left(\frac{1}{15}-\frac{1}{27}\,\mathrm{i}\right)    & \ket{0110101} & {}+\left(\frac{3}{77}-\frac{3}{61}\,\mathrm{i}\right)     & \ket{0110110} & {}-\left(\frac{5}{78}+\frac{5}{46}\,\mathrm{i}\right)     & \ket{0110111}     \\
				             {}+\left(\frac{3}{124}-\frac{6}{115}\,\mathrm{i}\right)         & \ket{0111000} & {}-\left(\frac{2}{65}-\frac{5}{59}\,\mathrm{i}\right)    & \ket{0111001} & {}-\left(\frac{3}{79}+\frac{3}{89}\,\mathrm{i}\right)     & \ket{0111010} & {}-\left(\frac{13}{113}+\frac{1}{30}\,\mathrm{i}\right)   & \ket{0111011}     \\
				             {}+\left(\frac{1}{9}-\frac{4}{111}\,\mathrm{i}\right)           & \ket{0111100} & {}+\left(\frac{2}{75}+\frac{9}{127}\,\mathrm{i}\right)   & \ket{0111101} & {}+\left(\frac{11}{129}+\frac{9}{140}\,\mathrm{i}\right)  & \ket{0111110} & {}+0                                                      & \ket{0111111}     \\
				             {}+\left(\frac{4}{45}+\frac{1}{95}\,\mathrm{i}\right)           & \ket{1000000} & {}+\left(\frac{7}{138}-\frac{3}{133}\,\mathrm{i}\right)  & \ket{1000001} & {}+\left(\frac{2}{83}+\frac{12}{113}\,\mathrm{i}\right)   & \ket{1000010} & {}-\left(\frac{4}{119}+\frac{9}{161}\,\mathrm{i}\right)   & \ket{1000011}     \\
				             {}+\left(\frac{1}{113}-\frac{4}{113}\,\mathrm{i}\right)         & \ket{1000100} & {}+\left(\frac{11}{145}+\frac{5}{158}\,\mathrm{i}\right) & \ket{1000101} & {}-\left(\frac{9}{103}-\frac{17}{190}\,\mathrm{i}\right)  & \ket{1000110} & {}+\left(\frac{2}{109}-\frac{13}{152}\,\mathrm{i}\right)  & \ket{1000111}     \\
				             {}+\left(\frac{6}{109}-\frac{3}{76}\,\mathrm{i}\right)          & \ket{1001000} & {}-\left(\frac{3}{125}-\frac{4}{49}\,\mathrm{i}\right)   & \ket{1001001} & {}+\left(\frac{3}{20}+\frac{7}{110}\,\mathrm{i}\right)    & \ket{1001010} & {}+\left(\frac{5}{159}+\frac{7}{163}\,\mathrm{i}\right)   & \ket{1001011}     \\
				             {}+\left(\frac{3}{46}-\frac{1}{46}\,\mathrm{i}\right)           & \ket{1001100} & {}+\left(\frac{1}{136}-\frac{1}{12}\,\mathrm{i}\right)   & \ket{1001101} & {}+\left(\frac{5}{177}-\frac{7}{128}\,\mathrm{i}\right)   & \ket{1001110} & {}+\left(\frac{1}{32}-\frac{3}{110}\,\mathrm{i}\right)    & \ket{1001111}     \\
				             {}+\left(\frac{1}{53}+\frac{8}{111}\,\mathrm{i}\right)          & \ket{1010000} & {}-\left(\frac{1}{152}-\frac{5}{56}\,\mathrm{i}\right)   & \ket{1010001} & {}+\frac{2}{131}                                          & \ket{1010010} & {}+\left(\frac{1}{58}-\frac{1}{31}\,\mathrm{i}\right)     & \ket{1010011}     \\
				             {}-\left(\frac{7}{61}+\frac{3}{53}\,\mathrm{i}\right)           & \ket{1010100} & {}+\left(\frac{1}{9}+\frac{3}{92}\,\mathrm{i}\right)     & \ket{1010101} & {}+\left(\frac{7}{183}-\frac{1}{25}\,\mathrm{i}\right)    & \ket{1010110} & {}+\left(\frac{7}{109}+\frac{5}{51}\,\mathrm{i}\right)    & \ket{1010111}     \\
				             {}-\left(\frac{10}{143}+\frac{22}{133}\,\mathrm{i}\right)       & \ket{1011000} & {}-\left(\frac{1}{16}+\frac{1}{60}\,\mathrm{i}\right)    & \ket{1011001} & {}-\left(\frac{3}{146}-\frac{3}{184}\,\mathrm{i}\right)   & \ket{1011010} & {}+\frac{14}{153}                                         & \ket{1011011}     \\
				             {}-\left(\frac{6}{173}+\frac{3}{149}\,\mathrm{i}\right)         & \ket{1011100} & {}+\left(\frac{9}{104}-\frac{3}{83}\,\mathrm{i}\right)   & \ket{1011101} & {}+\left(\frac{1}{8}+\frac{1}{85}\,\mathrm{i}\right)      & \ket{1011110} & {}+\left(\frac{1}{185}-\frac{5}{47}\,\mathrm{i}\right)    & \ket{1011111}     \\
				             {}-\left(\frac{5}{92}-\frac{7}{86}\,\mathrm{i}\right)           & \ket{1100000} & {}-\left(\frac{4}{93}-\frac{7}{86}\,\mathrm{i}\right)    & \ket{1100001} & {}-\left(\frac{5}{163}+\frac{1}{11}\,\mathrm{i}\right)    & \ket{1100010} & {}+\frac{1}{94}                                           & \ket{1100011}     \\
				             {}-\left(\frac{6}{131}-\frac{13}{149}\,\mathrm{i}\right)        & \ket{1100100} & {}+\left(\frac{5}{113}-\frac{7}{53}\,\mathrm{i}\right)   & \ket{1100101} & {}-\left(\frac{2}{69}-\frac{16}{151}\,\mathrm{i}\right)   & \ket{1100110} & {}-\left(\frac{1}{71}-\frac{5}{88}\,\mathrm{i}\right)     & \ket{1100111}     \\
				             {}-\left(\frac{4}{103}-\frac{1}{62}\,\mathrm{i}\right)          & \ket{1101000} & {}+\left(\frac{1}{101}-\frac{1}{63}\,\mathrm{i}\right)   & \ket{1101001} & {}+\left(\frac{1}{59}-\frac{2}{149}\,\mathrm{i}\right)    & \ket{1101010} & {}+\left(\frac{5}{137}-\frac{18}{149}\,\mathrm{i}\right)  & \ket{1101011}     \\
				             {}+\left(\frac{4}{43}+\frac{5}{117}\,\mathrm{i}\right)          & \ket{1101100} & {}+\left(\frac{5}{63}+\frac{1}{17}\,\mathrm{i}\right)    & \ket{1101101} & {}-\frac{5}{69}                                           & \ket{1101110} & {}-\left(\frac{3}{112}+\frac{5}{94}\,\mathrm{i}\right)    & \ket{1101111}     \\
				             {}+\left(\frac{1}{160}+\frac{3}{86}\,\mathrm{i}\right)          & \ket{1110000} & {}+\left(\frac{5}{104}-\frac{1}{25}\,\mathrm{i}\right)   & \ket{1110001} & {}+\left(\frac{1}{63}+\frac{6}{95}\,\mathrm{i}\right)     & \ket{1110010} & {}+\left(\frac{19}{109}-\frac{4}{51}\,\mathrm{i}\right)   & \ket{1110011}     \\
				             {}-\left(\frac{1}{36}+\frac{2}{107}\,\mathrm{i}\right)          & \ket{1110100} & {}+\left(\frac{1}{54}-\frac{5}{101}\,\mathrm{i}\right)   & \ket{1110101} & {}+\left(\frac{8}{159}+\frac{17}{164}\,\mathrm{i}\right)  & \ket{1110110} & {}+\left(\frac{1}{108}+\frac{7}{144}\,\mathrm{i}\right)   & \ket{1110111}     \\
				             {}-\left(\frac{1}{99}+\frac{6}{179}\,\mathrm{i}\right)          & \ket{1111000} & {}+\left(\frac{2}{127}+\frac{2}{55}\,\mathrm{i}\right)   & \ket{1111001} & {}-\left(\frac{5}{57}+\frac{1}{13}\,\mathrm{i}\right)     & \ket{1111010} & {}+\left(\frac{3}{44}-\frac{3}{79}\,\mathrm{i}\right)     & \ket{1111011}     \\
				             {}-\left(\frac{4}{67}+\frac{3}{94}\,\mathrm{i}\right)           & \ket{1111100} & {}-\left(\frac{7}{92}+\frac{5}{51}\,\mathrm{i}\right)    & \ket{1111101} & {}-\left(\frac{1}{13}+\frac{1}{73}\,\mathrm{i}\right)     & \ket{1111110} & {}+\left(\frac{10}{141}+\frac{1}{62}\,\mathrm{i}\right)   & \ket{1111111}     \\
				\end{array}
			$}
		\label{eq:expanded-second-seven-qubit-state}
	\end{equation}
	
	\newpage

	The seven-qubit TES states found are obtained from Haar-random seeds. In this sense, and despite the fact that TES states are not typical in the Hilbert space, TES states found in this way are generic in their set. To illustrate their structure, in Table~\ref{tab:TE7qubStats} we present the properties of state $|\tilde\phi\rangle$. Other examples are given in~\cite{Code-data}. This leads to TES states with $124$ nonzero real parameters, in comparison to the $2\cdot(2^7)-1 = 255$ free real parameters that seven-qubit states $\ket{\psi}\in(\C^2)^{\otimes 7}$ can have (namely $2^7$ complex parameters and one real constraint due to normalization).

	\begin{table}[h]
		\centering
		\setlength{\tabcolsep}{12pt} 
		\renewcommand{\arraystretch}{1.25} 
		\begin{tabular}{cccccccc}
			\hline
			Marginal  & Purity & $\theta$            & LHS   & RHS   & $\lambda_{\min}$ & $\lambda_{\max}$ & Rank \\
			\hline
			$(0,1,2)$ & 0.144  & -0.010              & 0.203 & 0.213 & 0.046            & 0.203            & 8    \\
			$(0,1,3)$ & 0.144  & -0.014              & 0.199 & 0.213 & 0.055            & 0.199            & 8    \\
			$(0,1,4)$ & 0.147  & $-1.1\cdot 10^{-4}$ & 0.192 & 0.192 & 0.039            & 0.192            & 8    \\
			$(0,1,5)$ & 0.145  & -0.005              & 0.210 & 0.215 & 0.059            & 0.210            & 8    \\
			$(0,1,6)$ & 0.146  & $-7.7\cdot10^{-6}$  & 0.202 & 0.202 & 0.041            & 0.202            & 8    \\
			$(0,2,3)$ & 0.146  & -0.005              & 0.201 & 0.206 & 0.056            & 0.201            & 8    \\
			$(0,2,4)$ & 0.145  & $-7.4\cdot 10^{-5}$ & 0.192 & 0.192 & 0.042            & 0.192            & 8    \\
			$(0,2,5)$ & 0.145  & $-4.5\cdot 10^{-4}$ & 0.200 & 0.200 & 0.045            & 0.200            & 8    \\
			$(0,2,6)$ & 0.148  & $-1.3\cdot10^{-5}$  & 0.210 & 0.211 & 0.058            & 0.210            & 8    \\
			$(0,3,4)$ & 0.144  & -0.011              & 0.190 & 0.200 & 0.049            & 0.190            & 8    \\
			$(0,3,5)$ & 0.143  & -0.028              & 0.195 & 0.223 & 0.056            & 0.195            & 8    \\
			$(0,3,6)$ & 0.139  & -0.044              & 0.189 & 0.233 & 0.050            & 0.189            & 8    \\
			$(0,4,5)$ & 0.148  & $-6.5\cdot 10^{-5}$ & 0.209 & 0.209 & 0.049            & 0.209            & 8    \\
			$(0,4,6)$ & 0.146  & $-2.6\cdot10^{-4}$  & 0.198 & 0.198 & 0.049            & 0.198            & 8    \\
			$(0,5,6)$ & 0.143  & -0.011              & 0.214 & 0.224 & 0.053            & 0.214            & 8    \\
			$(1,2,3)$ & 0.146  & -0.010              & 0.198 & 0.208 & 0.043            & 0.198            & 8    \\
			$(1,2,4)$ & 0.146  & $-1.1\cdot10^{-4}$  & 0.194 & 0.194 & 0.047            & 0.194            & 8    \\
			$(1,2,5)$ & 0.148  & $-4.3\cdot10^{-4}$  & 0.191 & 0.191 & 0.052            & 0.191            & 8    \\
			$(1,2,6)$ & 0.144  & $-2.5\cdot10^{-4}$  & 0.197 & 0.197 & 0.038            & 0.197            & 8    \\
			$(1,3,4)$ & 0.148  & -0.004              & 0.203 & 0.207 & 0.049            & 0.203            & 8    \\
			$(1,3,5)$ & 0.145  & -0.002              & 0.195 & 0.197 & 0.043            & 0.195            & 8    \\
			$(1,3,6)$ & 0.141  & -0.028              & 0.192 & 0.220 & 0.046            & 0.192            & 8    \\
			$(1,4,5)$ & 0.145  & $-3.0\cdot10^{-4}$  & 0.202 & 0.202 & 0.049            & 0.202            & 8    \\
			$(1,4,6)$ & 0.148  & -0.001              & 0.203 & 0.203 & 0.048            & 0.203            & 8    \\
			$(1,5,6)$ & 0.145  & -0.002              & 0.192 & 0.195 & 0.035            & 0.192            & 8    \\
			$(2,3,4)$ & 0.144  & -0.001              & 0.206 & 0.207 & 0.042            & 0.206            & 8    \\
			$(2,3,5)$ & 0.143  & -0.011              & 0.195 & 0.205 & 0.047            & 0.195            & 8    \\
			$(2,3,6)$ & 0.145  & -0.011              & 0.198 & 0.209 & 0.056            & 0.198            & 8    \\
			$(2,4,5)$ & 0.146  & $-2.5\cdot10^{-4}$  & 0.197 & 0.198 & 0.042            & 0.197            & 8    \\
			$(2,4,6)$ & 0.148  & $-1.3\cdot10^{-4}$  & 0.195 & 0.196 & 0.051            & 0.195            & 8    \\
			$(2,5,6)$ & 0.142  & -0.029              & 0.191 & 0.220 & 0.060            & 0.191            & 8    \\
			$(3,4,5)$ & 0.145  & $-3.6\cdot10^{-4}$  & 0.196 & 0.197 & 0.047            & 0.196            & 8    \\
			$(3,4,6)$ & 0.146  & $-6.5\cdot10^{-5}$  & 0.189 & 0.189 & 0.045            & 0.189            & 8    \\
			$(3,5,6)$ & 0.140  & -0.033              & 0.192 & 0.224 & 0.059            & 0.192            & 8    \\
			$(4,5,6)$ & 0.148  & $-7.6\cdot10^{-6}$  & 0.198 & 0.198 & 0.053            & 0.198            & 8    \\
			\hline
		\end{tabular}
		\caption{{\bf Properties of the 35 three-qubit marginals of the analytical 7-qubit TES state~\eqref{eq:expanded-second-seven-qubit-state}}, found by minimizing Eq.~\eqref{eq:CostFunctionAS} with an initial Haar-random seed. The columns correspond respectively to the marginal considered $\varrho_{ij\ell}$ with $(i,j,\ell)$, the purity $\tr(\varrho_{ij\ell}^2)$, the cost function $\theta=\mathrm{LHS}-\mathrm{RHS}$ (where $\text{LHS}=\lambda_{1}^{S}$ and $\text{RHS}=\lambda_{D-1}^{S}+2\sqrt{\lambda_{D-2}^{S}\lambda_{D}^{S}}$ according to Eq.~\eqref{eq:AShildebrand}), minimal and maximal eigenvalues of the marginal, and its rank. The reductions are absolutely separable between each qubit and the rest if $\theta\leq 0$.}
		\label{tab:TE7qubStats}
	\end{table}

\section{Error correcting bounds and proof of Theorem \ref{th:lb}}\label{app:minpurity}
	In this section, we explain the mathematical machinery used to determine the non-existence of quantum error-correcting codes and its connection to finding the minimal average purity as defined in Eq.~\eqref{eq:LBpurity}. This will prove Theorem~\ref{th:lb}.

	A quantum error-correcting code with parameters $(\!(n,K,\delta)\!)_2$ encodes a $K$-dimensional Hilbert space to a code subspace of $n$-qubit Hilbert space. The parameter {\em distance} $\delta$ denotes that the code can detect errors acting nontrivially on at most $\delta-1$ qubits.
	An encoding operation is a projection of the Hilbert space onto the code subspace, and therefore each code is uniquely defined by its associated projector $\Pi$.
	A fundamental question is to determine whether the existence of a qubit quantum code (or projector $\Pi$) with parameters $(\!(n,K,\delta)\!)_2$ is possible
	~\cite{PhysRevA.55.900,PhysRevLett.78.1600,681315}.
	To tackle this problem, linear programming (LP) bounds~\cite{PhysRevLett.78.1600,681315,681316,796376} have been derived, and recent improvements have been achieved with SDP bounds~\cite{munne2025sdpboundsquantumcodes,munne2026sdpboundsquantumcodes}.
	In particular, using the LP for quantum codes one can approximate the sum of purities of the marginals of $\Pi$ for a given set $S$ of size $j$~\cite{681316},
	\begin{align}
		A'_j= \sum_{S:|S|= j} \tr\big ((\tr_{S^c}(\Pi)^2 \big ) .
	\end{align}
	Since we are interested in properties of pure states (rank-one projectors), we will focus on quantum codes with parameters $(\!(n,1,\lfloor n/2\rfloor+1)\!)_{2}$~\cite{681315}, whose code projector $\Pi=\ketbra{\psi}{\psi}$ is one-dimensional. These cannot be used to encode information, but have the special property that $\ket{\psi}$ is an $n$-qubit AME state~\cite{Scott2004QECentPower}.
	This is because $(\!(n,1,\lfloor n/2\rfloor+1)\!)_{2}$-quantum codes also known as \emph{self-dual quantum codes} are defined by convention to be pure codes, i.e., $\bra{\psi}E\ket{\psi}=0$ for all errors $E$ acting nontrivially on $\lfloor n/2\rfloor$ or fewer qubits.
	More precisely, the existence of an $(\!(n,1,\lfloor n/2\rfloor+1)\!)_{2}$ quantum code determines the existence of a corresponding $n$-qubit AME state and vice-versa.
	Using this correspondence, the nonexistence of AME states has been addressed as a feasibility LP with the following constraints: any rank-one projector $\Pi=\dyad{\psi}$ must satisfy $A'_0=1$ due to normalization, $A'_{n-j}=A'_{j}$ since the purity of any marginal $\varrho_S$ is equal to that of its complement $\varrho_{S^c}$, and moreover the inequality
	\begin{equation}
		A_j:=\sum^j_{\ell=0} (-1)^{j-\ell}2^{\ell} \binom{n-\ell}{n-j}A'_\ell\geq 0 \,
	\end{equation}
	for all subset sizes $j=0,\dots, n$, where $A_j$ is one of the so-called {\em quantum weight enumerators}. This last condition follows by averaging over all subsystems of size $j$ the result of~\cite[Theorem 15]{681316}.
	The values of the marginal purities are further restricted by the so-called {\em shadow enumerators}~\cite{796376,1751-8121-51-17-175301},
	\begin{align}
		S_j:=\sum^n_{\ell=0} K_{n-j}(\ell) A'_\ell ,
	\end{align}
	where $K_u(x)=\sum^{n}_{\alpha=0} \binom{n-x}{u-\alpha}\binom{x}{\alpha}(-1)^\alpha$ are the \emph{Krawtchouk polynomials}.
	These enumerators further satisfy $S_j=0$ when $n-j$ is odd and $S_j\geq 0$ when $n-j$ is even.
	If $\ket{\psi}$ is AME, then the purity is constrained by
	$A'_j=\binom{n}{j} 2^{-j}$ for $i=1,\dots, \lfloor n/2\rfloor$, which is equivalent to the pure code condition.
	Using these relations, we lower bound Eq~\eqref{eq:LBpurity} as $\hat{p}_{\text{LB}} \leq p_{\text{LB}}$ with the optimization program of Theorem~\ref{th:lb}, where the variables $a'_j$ relax marginal purities $A'_j$.

	\section{Maximal purity of absolutely separable states in $\mathbb{C}^{2}\otimes\mathbb{C}^{d}$ and proof of Theorem~\ref{th:ub}}
	\label{App:ProofMaxPurityAS2d}
    In this section, we show how to bound on the maximal purity $\sum_{i=1}^{2d}\lambda_{i}^{2}$ of $\text{AS}_{1|r-1}$ states in $\mathbb{C}^{2}\otimes\mathbb{C}^{d}$ is obtained.

	First, note that the $\text{AS}_{1|r-1}$ condition for the non-increasingly ordered eigenvalues $\{\lambda_{i}^{\downarrow}\}$ only depends on $4$ out of the $2d$ eigenvalues:
	\begin{equation}
		\label{eq:AppendixAS}
		\lambda_{1}\leq \lambda_{2d-1}+2\sqrt{\lambda_{2d-2}\lambda_{2d}}.
	\end{equation}

	Thus, we can consider that $\lambda_{1}\geq \lambda_{i}\geq\lambda_{2d-2}$ for all $i\in[1,2d-2]$. Therefore, the quantity $(\lambda_{i}-\lambda_{1})(\lambda_{i}-\lambda_{2d-2})\leq 0$. Isolating $\lambda_{i}^{2}\leq (\lambda_{1}+\lambda_{2d-2})\lambda_i-\lambda_1\lambda_{2d-2}$ from the previous equation, allows us to compute:
	\begin{equation}
		\label{eq:Appendix(1)Proof}
		\sum_{i=1}^{2d-2}\lambda_{i}^{2}\leq \left(\lambda_{1}+\lambda_{2d-2}\right)\sum_{i=1}^{2d-2}\lambda_i-(2d-2)\lambda_1\lambda_{2d-2}.
	\end{equation}

	Next, considering the ordered eigenvalues, $\lambda_{i}^{2}\leq \lambda_{i-1}\lambda_i$ and the normalization condition $\sum_{i}\lambda_i=1$, we can bound:
	\begin{equation}
		\label{eq:Appendix(2)Proof}
		\sum_{i=1}^{2d}\lambda_{i}^{2}\leq \lambda_{1}+\lambda_{2d-2}-(2d-2)\lambda_{1}\lambda_{2d-2}-\lambda_{1}(\lambda_{2d-1}+\lambda_{2d}).
	\end{equation}
	We can keep reexpressing the upper bound through the algebraic identity $2\sqrt{\lambda_{2d-2}\lambda_{2d}}\leq \lambda_{2d-2}+\lambda_{2d}$ and Eq.~\eqref{eq:AppendixAS}, bounding the last summand of Eq.~\eqref{eq:Appendix(2)Proof} as:
	\begin{equation}
		\label{eq:AppendixExtra}
		\sum_{i=1}^{2d} \lambda_{i}^{2}\leq \lambda_{1}+\lambda_{2d-2}-\lambda_{1}^{2}-(2d-3)\lambda_{1}\lambda_{2d-2}.
	\end{equation}

	Finally, we consider also the ratio $x=\lambda_{1}/\lambda_{2d-2}$, which can be bounded from Eq.~\eqref{eq:AppendixAS} as $1\leq x \leq 3$ to obtain:
	\begin{equation}
		\sum_{i=1}^{2d}\lambda_{i}^{2}\leq (x+1)\lambda_{2d-2}-x(2d-3+x)\lambda_{2d-2}^{2}.
	\end{equation}
	It is now possible to maximize over $\lambda_{2d-2}$ the previous equation to obtain
	\begin{equation}
		\label{eq:AppendixAnsatz}
		\sum_{i=1}^{2d}\lambda_{i}^{2}\leq \frac{(x+1)^{2}}{4x(2d-3+x)}.
	\end{equation}

	We need to study the behaviour of the function over $x$. It is possible to check that it has a minimum at $x^{*}=\frac{2d-3}{2d-5}$ and it increases over the bounds of $x\in[1,3]$. Therefore, the maximal values have to be on one of the extremes. If we analyze $f(1)=1/(2d-2)$ and $f(3)=4/(3\cdot 2d)$, we realize that for $d > 4$, the maximal value of the function over the valid values of $x$ is given by:
	\begin{equation}
		\label{eq:AppendixPurity}
		\sum_{i=1}^{2d}\lambda_{i}^{2}\leq \frac{4}{3\cdot 2d}.
	\end{equation}
	Moreover, this upper bound is, indeed, attainable for even values of $d$ by the vectors of eigenvalues of the form:
	\begin{equation}
		\label{eq:AppendixAttainability}
		\left(\underbrace{\frac{1}{d}, \cdots, \frac{1}{d}}_{d/2}, \underbrace{\frac{1}{3d}, \cdots , \frac{1}{3d}}_{3\cdot d/2}\right).
	\end{equation}

	This proves Conjecture 3.1. of  \cite{dung_maximum_2025} and Conjecture 6.7. of \cite{ahiable_geometry_2026} for the qubit-qudit case. An alternative proof, through an exhaustive study of the extreme points of the set can also be found in Ref.~\cite{Song_2025_Extreme_APPT}.

	\section{Separability structure of the absolutely separable marginals} \label{Appendix:separability_structure}

	In this appendix we address the question of why to consider the reductions of the TES states as being $\text{AS}_{1|r-1}$ in the partition $\mathbb{C}^{2}\otimes\mathbb{C}^{2^{r-1}}$. We also introduce Conjecture~\ref{Conj:MaximalPurityAS4d} for the non-existance of pure states with fully absolutely separable $\lfloor n/2 \rfloor$-qubit marginals besides $n=7,9$. Thus, posing $\text{AS}_{1|r-1}$ in the partition $\mathbb{C}^{2}\otimes\mathbb{C}^{2^{r-1}}$ as the maximal attainable unitary invariant separability for all the $\lfloor n/2 \rfloor$-qubit marginals.\\ 

	First, we consider the set of absolutely positive-partial-transpose ($\text{APPT}_{k|r-k}$) states as a relaxation of absolute separability. Notice that, since APPT and AS are global unitary invariant properties, they can be certified solely from the spectra $\{\lambda_{i}\} $ of the state under scrutiny. Thus, for APPT detection it is enough to consider the state as the vector of its eigenvalues. The set of $\text{APPT}_{k|r-k}$ states has been completely characterized through a family of linear matrix inequalities (LMIs)~\cite{hildebrand_ppt_2007}. Although $\text{APPT}_{k|r-k}$ and $\text{AS}_{k|r-k}$ are conjectured to coincide in arbitrary dimensions, their equivalence has so far only been established for qubit-qudit systems~\cite{arunachalam_is_2015,johnston_separability_2013}. The condition that a $r$-qubit marginal of a pure state is $\text{APPT}_{1|r-1}$ with respect to the bipartition $\mathbb{C}^{2}\otimes\mathbb{C}^{2^{r-1}}$  (see Eq.~\eqref{eq:AppendixAS}) is weaker than the corresponding condition for $\text{APPT}_{2|r-2}$ in the bipartition $\mathbb{C}^{4}\otimes\mathbb{C}^{2^{r-2}}$. Indeed, the LMI characterizing the former case appears as the first $2\times2$ principal minor of one of the LMI associated with the latter. Consequently, $\text{APPT}_{2|r-2}$ imposes all the constraints arising from the $\text{APPT}_{1|r-1}$ qubit bipartition, together with plenty of additional ones. The same fact persists as the bipartition becomes increasingly balanced. Thus, requiring absolute separability across more balanced partitions leads to progressively smaller spectral sets completely contained in the less balanced ones. \\

	Second, notice that a $r$-qubit system can be considered in the bipartitions from $1|r-1$ up to the most balanced one $\lfloor r/2\rfloor|\lceil r/2 \rceil$. For the cases up to $n=7$, the reductions to $\lfloor n/2 \rfloor$ qubits that we consider in the main text have up to $3$ qubits, thus only allowing one possible bipartition $1|1$ or $1|2$. The first case that allows for different bipartitions is $n=8$, which we have shown that does not exist. Next, we study $\text{APPT}_{2|r-2}$ in the partition $\mathbb{C}^{4}\otimes\mathbb{C}^{2^{r-2}}$ to have a deeper insight for $n\geq 8$. For this case, we have not managed to prove the maximal purity of the set. Nevertheless, we conjecture a maximal value for it that coincides also with other conjectures in the literature~\cite{ahiable_geometry_2026}.

	\begin{proposition}
		\label{Prop:MaxPurityOfAPPT_2evals}
		Let $\varrho\in\mathbb{C}^{2^{2}}\otimes\mathbb{C}^{2^{r-2}}$ be an absolutely PPT state with only two distinct eigenvalues. Its maximal purity is given by:
		\begin{equation}
			\max_{\lambda\in APPT}\sum_{i}\tr(\varrho^{2})=
			\begin{cases}
				\dfrac{2^r+8}{\left(2^r+2\right)^2},
				                     & r=4,5, \quad \text{attained by } \frac{1}{2^r+2}\left(3,1,\cdots,1\right).\\[8pt]
				\dfrac{2^{4-r}}{15}, & r\geq 6, \quad \text{attained by } \left(\underbrace{\frac{1}{3\cdot2^{r-2}}, \cdots, \frac{1}{3\cdot2^{r-2}}}_{3\cdot2^{r-2}/2}, \underbrace{\frac{1}{5\cdot2^{r-2}}, \cdots , \frac{1}{5\cdot2^{r-2}}}_{5\cdot2^{r-2}/2}\right).
			\end{cases}
		\end{equation}
	\end{proposition}
	\begin{proof}
		For this case, the $\text{APPT}_{2|r-2}$ condition is equivalent to the positivity of twelve $4\times4$ LMI \cite{hildebrand_ppt_2007}. It is possible to consider the maximal purity restricted only to the first of the LMIs. Notice that computing the maximal purity for just one LMI sets an upper bound for the maximal purity of APPT, since the rest of the matrices will only add more constraints to the vectors of eigenvalues.

		\begin{equation}
			\label{eq:FirstLMI4d}
			M=\begin{pmatrix}
				2\lambda_{4d}            & \lambda_{4d-1}-\lambda_1 & \lambda_{4d-2}-\lambda_2 & \lambda_{4d-3}-\lambda_3 \\
				\lambda_{4d-1}-\lambda_1 & 2\lambda_{4d-4}          & \lambda_{4d-5}-\lambda_4 & \lambda_{4d-6}-\lambda_5 \\
				\lambda_{4d-2}-\lambda_2 & \lambda_{4d-5}-\lambda_4 & 2\lambda_{4d-7}          & \lambda_{4d-8}-\lambda_6 \\
				\lambda_{4d-3}-\lambda_3 & \lambda_{4d-6}-\lambda_5 & \lambda_{4d-8}-\lambda_6 & 2\lambda_{4d-9}
			\end{pmatrix}\geq0
		\end{equation}

		It is straightforward to reproduce the proof of Appendix~\ref{App:ProofMaxPurityAS2d} considering only states with two-level spectra: $(\underbrace{a,\cdots,a}_{\ell},b,\cdots,b)$ and maximizing the purity given Eq.~\eqref{eq:FirstLMI4d}. As it can be seen, different values of $\ell$ lead to different maximal values of the purity that need to be compared. The proposed value is not only the maximal purity among the two-level spectra, but also among all the spectra majorized by any two-level spectra due to strict Schur convexity of the purity~\cite{marshall_inequalities_2011}.
	\end{proof}

	The computed result, nevertheless, does not consider all possible APPT states. We conjecture that the maximal purity of the $\text{APPT}_{2|r-2}$ set is indeed obtained by a state with only two distinct eigenvalues.
	\begin{conjecture}
		\label{Conj:MaximalPurityAS4d}
		The maximal purity of an APPT state in $\mathbb{C}^{2^{2}}\otimes\mathbb{C}^{2^{r-2}}$ is attained by a two-level spectrum.
	\end{conjecture}

	\textit{Sketch of the proposed conjecture.} (See also Conjecture 6.7. of \cite{ahiable_geometry_2026})
	Numerical maximizations support our results, but we have not been able to proof that the conjectured results are the absolute maximum over the set. It can also be seen that the proposed maximizing spectra for the two-level spectra saturates diagonal dominance of the different LMI, being the states with maximal purity to do so. Nevertheless, diagonal dominance is only a sufficient condition for positive semidefiniteness of the LMI, thus allowing for the existance of states with higher purity.

	Given strict Schur convexity of the purity, we know that the maximizer should be an extreme point of the $\text{APPT}_{2|r-2}$ set. Nevertheless, it is also known that the set has infinitely many extreme points, and most of them remain uncharacterized up to date.

	Finally, defining the gaps $\delta_i:=\lambda_i-\lambda_{i+1}\geq0$ also leads to a possible approach to complete the proof. Proving that the maximizer spectra only has one gap different than $0$ or showing that all $\text{APPT}_{2|r-2}$ states with more than one non-zero gap are always majorized by a state with one non-zero gap would verify the conjecture. \qed

    \vspace{0.3cm}
	When we recreate Figure~\ref{fig:Puritycombined} with the conjectured bound for the next separability case, we see that TES states with $\text{AS}_{2|r-2}$ marginals in any partition $\mathbb{C}^{{2}^{2}}\otimes\mathbb{C}^{2^{r-2}}$ can only exist for $n=9$ qubits.

	\begin{figure}
		\centering
		\begin{minipage}{0.48\linewidth}
			\centering
			\includegraphics[width=\linewidth]{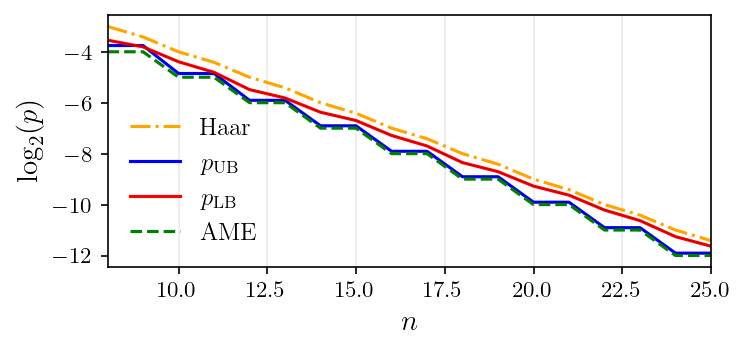}
		\end{minipage}
		\hfill
		\begin{minipage}{0.48\linewidth}
			\centering
			\includegraphics[width=\linewidth]{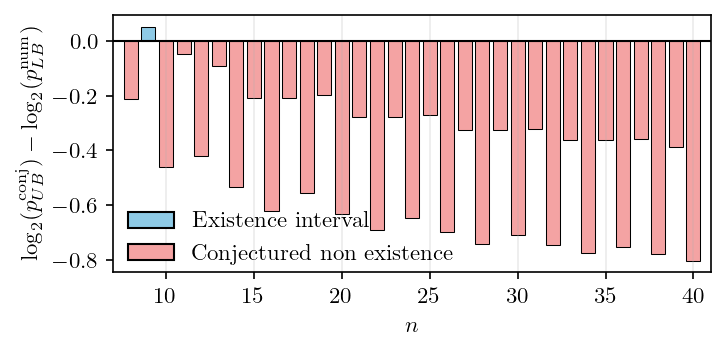}
		\end{minipage}

		\caption{\textbf{Existence of TES states between $8$ and $40$ qubits.}
			\textit{Left:} We compare the lower bounds found numerically in this
			work with the LP from Eq.~\eqref{eq:lp} (Approach 1), and the upper
			bounds on the maximal purity given by
			Conjecture~\ref{Conj:MaximalPurityAS4d}.}
		\label{fig:4xn_double}
	\end{figure}

	Finally, one could consider yet another separability structure~\cite{dur_classification_2000, guhne_entanglement_2009}: full absolute separability. This refers to a state that can be written as a combination of the tensor product of their parts, and that the state remains fully separable under the action of any global unitary. At present, however, the set of absolutely fully separable states has not been characterized beyond known separable balls around the maximally mixed state~\cite{gurvits_better_2005,hildebrand_entangled_2007}, which are purity-based criteria.  Moreover, absolute full separability is known to be substantially more restrictive than $\text{APPT}_{\lfloor\frac{r}{2}\rfloor | \lceil\frac{r}{2}\rceil}$. Next, we are going to briefly study full absolute separability for the only possible cases $n=7,9$, since they are the qubit systems for which fully-AS marginals could exist according to Fig.~\ref{fig:4xn_double}.\\

	We recall that if $\hat{p}_{LB}>p_{UB}$ then TES states cannot exist. All $3$-qubit states with purity under $\approx0.140$ \cite{hildebrand_entangled_2007} will be fully-AS, and for the $4$-qubit ones, the bound is at $\approx0.065$ \cite{hildebrand_entangled_2007}. Even though this bounds are only sufficient conditions for full-AS, the comparison with the bounds on the lower bound given by Eq.~\eqref{eq:lp}, namely $\hat{p}_{LB}\approx0.127$ and $\hat{p}_{LB}\approx0.071$ is not promising since for the case of $9$-qubit systems, no TES state with fully-AS marginals can be certified with the state of the art knowledge. For the $7$-qubit case, despite our efforts, we have not been able to certify any TES state with all the reduced marginals being fully-AS, but we have found pure states that have some of the reduced marginals Fully-AS (see the (0,3,6) marginal in Table~\ref{tab:TE7qubStats}). For the case of $n=9$ qubits, moreover, we have not been able to find any example of TES state for any AS structure of the marginals. Moreover, the short existance interval displayed in Figure~\ref{fig:4xn_double} also seems to suggest non-existance of TES states with fully-AS marginals.
    Thus, the separability structure of marginals in AS of $\mathbb{C}^{2}\otimes\mathbb{C}^{2^{r-1}}$, being the less restrictive of all the AS sets, seems to be the only attainable case in general. Thus, is the most separable structure that all the $\lfloor n/2\rfloor$-qubit marginals can have.\\

	By studying the separability structure of the marginals of potential TES states, we have provided also valuable insight on the general marginal problem.

	\newpage

	\section{Increase of magic when maximizing entanglement}\label{app:Magic}
Here we will analyze the entanglement and magic induced in states within the gradient descent algorithm. We initialize our optimization algorithm with Haar-random states on $n$-qubits which minimizes the cost function as defined in \eqref{eq:CostFunctionAS} As summarized in Figure~\ref{tab:MagicPurity}, initializing with $2000$ Haar-random states yields $1995$ TES for $n=7$ and $2000$ TES for $n=4$. Both for $n=4$ and $n=7$, the resulting TES contains a substantially large amount of magic, which is comparable to the typical values for Haar-random states. This naturally raises a fundamental question: Is the high magic content in these TES inherited from the initial seed states, or is it generated dynamically during the optimization process?

To isolate the contribution of initial magic, we contrast these high-magic Haar-random seeds with initial states that possess strictly \emph{zero} magic. Specifically, stabilizer states. A stabilizer state $\ket{\psi}$ on $n$-qubits is the unique state in the eigenspace with eigenvalue $+1$ of an $n$-generator stabilizer group $\mathcal{S}$ containing $2^n$ Pauli strings. For any stabilizer state $\ket{\psi}$, the Pauli overlap in Eq.~\eqref{eq:MagicRenyi} satisfies:
\begin{align}
|\braket{\psi}{P|\psi}|^2 = \begin{cases} 1, & \forall P \in \mathcal{S}, \\ 0, & \forall P \notin \mathcal{S}. \end{cases}
\end{align}

Consequently, for stabilizer states, the Stabilizer Rényi Entropy vanishes for all $\alpha$, $S_\alpha(\ket{\psi}) = 0$.
We use the numerical routine in Ref.~\cite{qotlabs_randstab} to generate uniformly random stabilizer states to seed our gradient descent optimization algorithms. Initializing $2000$ random stabilizer states yields $71$ TES for $4$-qubit stabilizer states after $200$ optimization steps and $207$ TES for $7$-qubit stabilizer states after $2000$ optimization steps. Crucially, despite starting in this way from states with \emph{zero} magic, the optimization still converges to TES with large magic. In contrast, trajectories that fail to reach a TES (non-TES) do not exhibit this drastic increase in magic (See Figure~\ref{tab:MagicPurity}).
\\

\textbf{Interpolating Initial Magic using the Stabilizer Gap}: These observations raise the question of whether the magic of the initial states (besides the extremal cases of 0 magic (for stabilizer random states) and high magic (for Haar-random states). To systematically investigate the regime in between, we generate initial states in three steps:
\begin{enumerate}
\item Generate an $n$-qubit Haar-random state $\ket{\phi}$.
\item Construct a projector onto an $(n-q)$-qubit random stabilizer state $\ket{\psi}_{n-q}$.
\item Project $\ket{\phi}$ onto the $n$-qubit space using the operator:
\begin{equation*}
    \mathbb{P} = \ketbra{\psi}_{n-q} \otimes \mathbb{I}_q
\end{equation*}
\end{enumerate}

Where $\mathbb{I}_q$ denotes the identity operator on the remaining $q$ qubits. Hence, we define the integer $q \in \{0, 1, \dots, n\}$ as the stabilizer gap, representing the number of unconstrained qubit degrees of freedom. We can build the intuition about the stabilizer gaps as follows: $q = 0$ the state is projected onto an $n$-qubit stabilizer state. Hence, yielding pure stabilizer seed states with zero initial magic ($S_2 = 0$) and when, $q = n$ the projector reduces to the identity operator $\mathbb{P} = \mathbb{I}_n$, recovering the unconstrained $n$-qubit Haar-random seeds. Finally, when $0 < q < n$, only $n-q$-qubits of the full $n$-qubits are constrained to be a stabilizer.
This variation in initial magic across stabilizer gaps $q$ can be seen in the green in the top panels of Figs.~\ref{fig:4q_4000_Hist} and \ref{fig:7q_2000_Hist}, bounded by the extremes $q = 0$ and $q = n$.

\textbf{Dynamics of Magic and subsystem-purity}:
Running the optimization algorithm on states across different stabilizer gaps $q$ reveals that, upon filtering by TE-ness, the final magic ($S_2$) for states that become TES is significantly larger than for non-TES states. Importantly, this feature persists regardless of the amount of initial magic of the seed states.

To understand the mechanism underlying this increase in magic, we track the change of magic alongside subsystem purity (a metric closely linked to total entanglement) across optimization steps. These dynamics are illustrated in Figs.~\ref{fig:7q_2000_TE} and \ref{fig:4q_4000_TE}. The left subplots show the mean of the average purity of the smaller subsystem evaluated over all the largest bipartitions. For example, in the $7$-qubit system, we consider all $3\vert{}4$ qubit bipartitions and take the average purity over the $3$-qubit subsystems. The central subplots show the maximum subsystem purity across all the bipartitions. \emph{Note}: If this maximum purity drops below the analytical bound established by Gurvits \cite{gurvits_largest_2002}, the state is necessarily a TES and the right subplots show the full-system Stabilizer $2$-Rényi Entropy $S_2$.

A contrast emerges when comparing the trajectory of TES against non-TES states (see Figs.~\ref{fig:4q_4000_NTE} and \ref{fig:7q_2000_NTE}). For states that evolve into TES, the decrease in subsystem purity (in both max purity and average purity) directly coincides with a steady increase of magic over the course of optimization. Instead, in non-TES trajectories, this coupled dynamical behavior is not as prominent.

The bottom subplots of Figs.~\ref{fig:4q_4000_Hist} and \ref{fig:7q_2000_Hist} summarize the final proportion of states identified as TES versus non-TES across all runs.

Although our state searches were extensive, we note that specific stabilizer gaps. Specifically, $q = n - 1$ yields a significantly lower count of TES compared to the other stabilizer gaps (below we specifically compare the counts to $q=2$) within the same fixed number of optimization steps. For $n = 4$ qubits with a stabilizer gap of $q = 3$, out of $4000$ initial seeds, only $20$ TES were obtained (compared to $3797$ TES for $q = 4$). Similarly, for $n = 7$ qubits with a stabilizer gap of $q = 6$, out of $2000$ initial seeds, only $5$ TES were found (compared to $2000$ TES for $q = 7$).

	\newpage
	\begin{figure}[p] 
		\centering
		\includegraphics[width=0.75\textwidth]{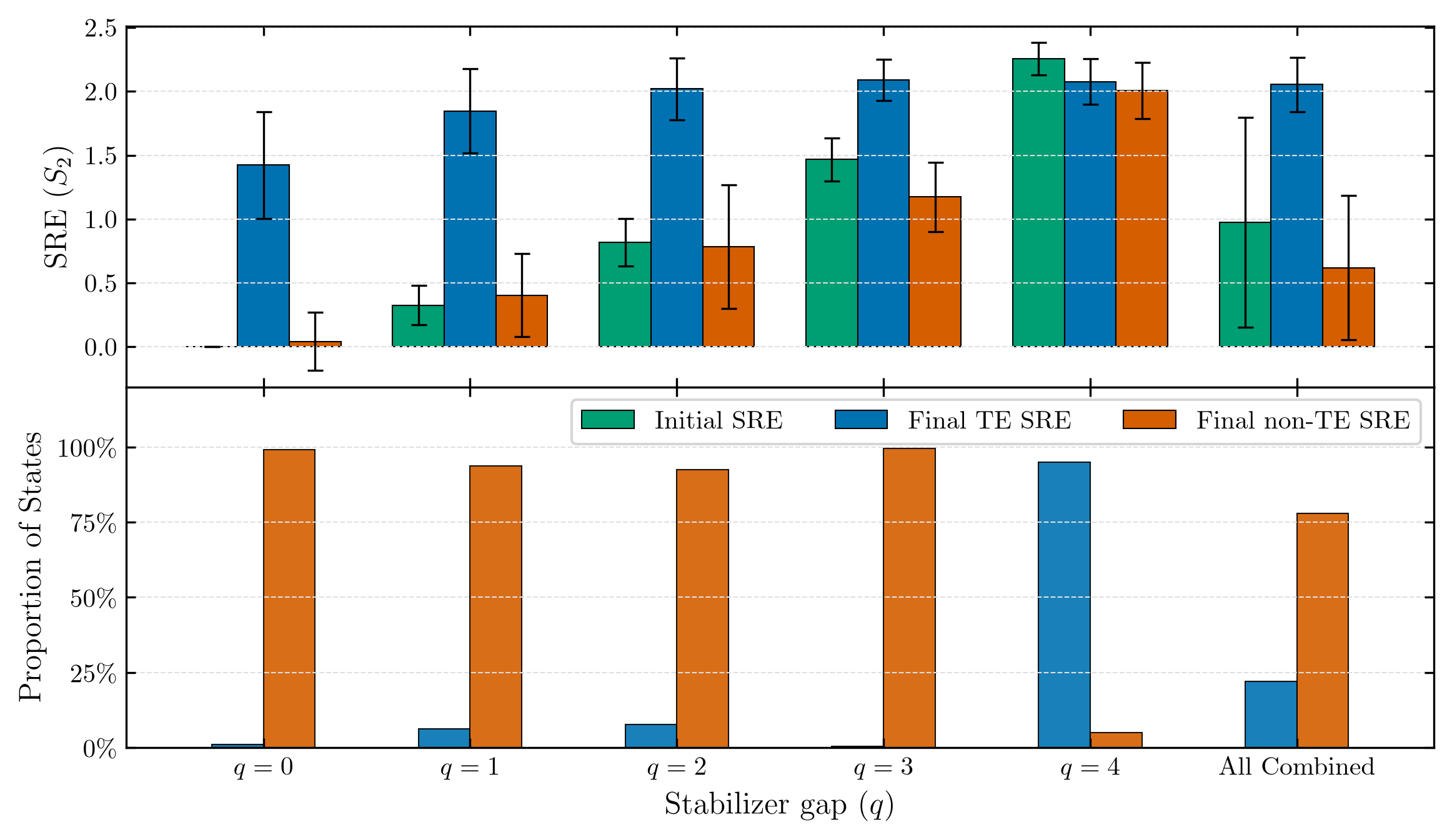}
		\caption{Top: comparison of SRE of the initial $4000$ seed of $4$-qubit states and states obtained after the $100$-step gradient descent optimization separated by TE-ness of the final states. Bottom: Proportion of states that ended up being TES or Non-TES  }
		\label{fig:4q_4000_Hist}
		\vspace{0.5cm} 
		\includegraphics[width=0.90\textwidth]{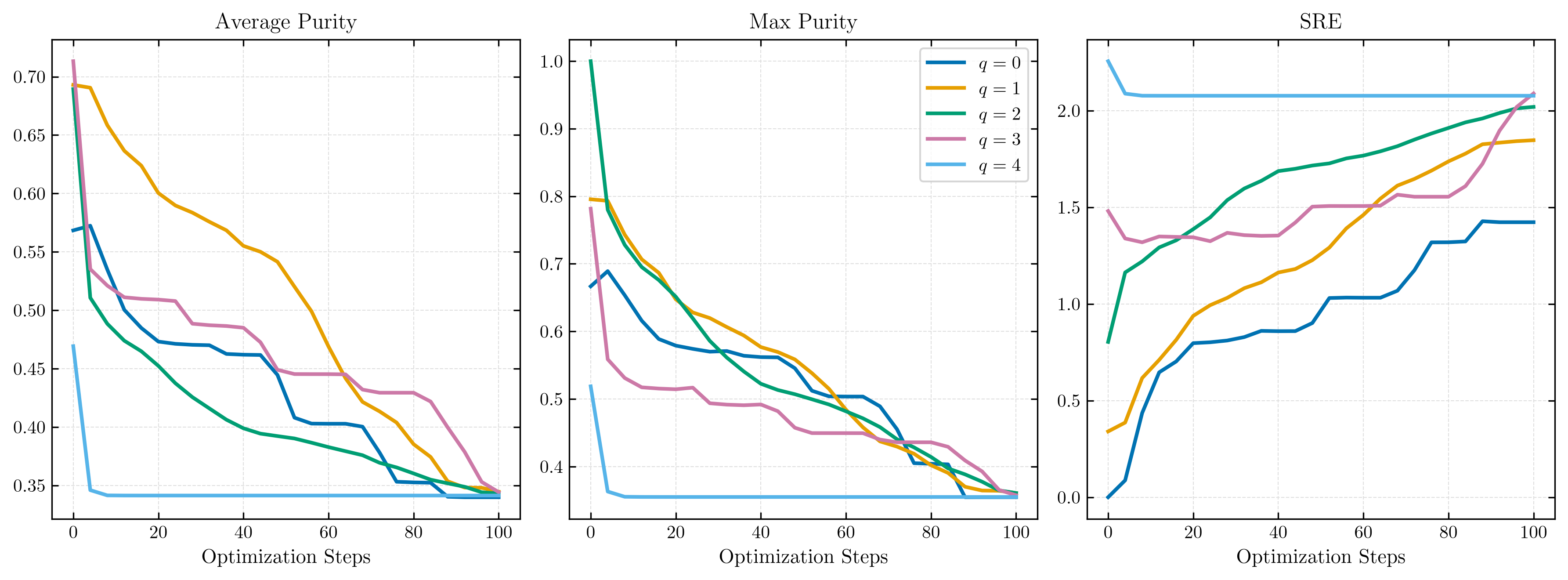}

		\caption{Average Purity, Maximal Purity and SRE for $4$-qubit seed-states that remain TES after $100$ steps of gradient descent optimization.}
		\label{fig:4q_4000_TE}
		\vspace{0.5cm}
		\includegraphics[width=0.90\textwidth]{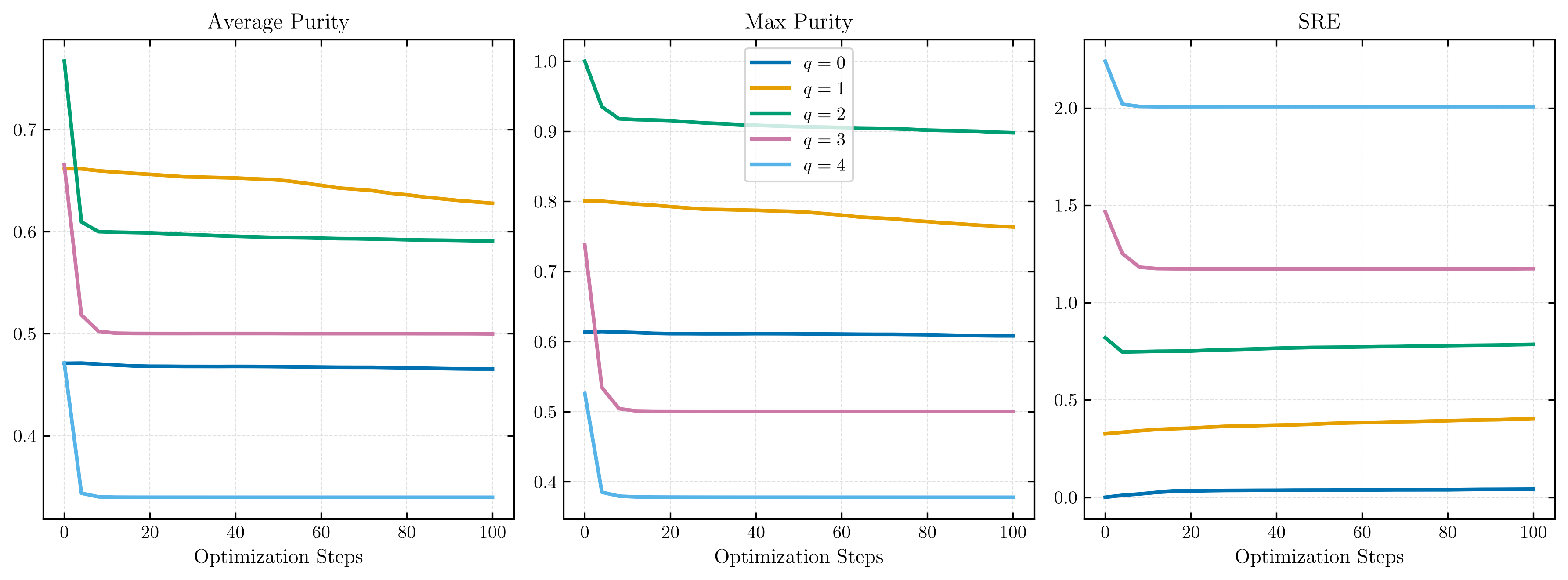}
		\caption{Average Purity, Maximal Purity and SRE for $4$-qubit seed-states that remain non-TES after $100$ steps of gradient descent optimization.}
		\label{fig:4q_4000_NTE}
	\end{figure}
	\begin{figure}[p] 
		\centering
		\includegraphics[width=0.75\textwidth]{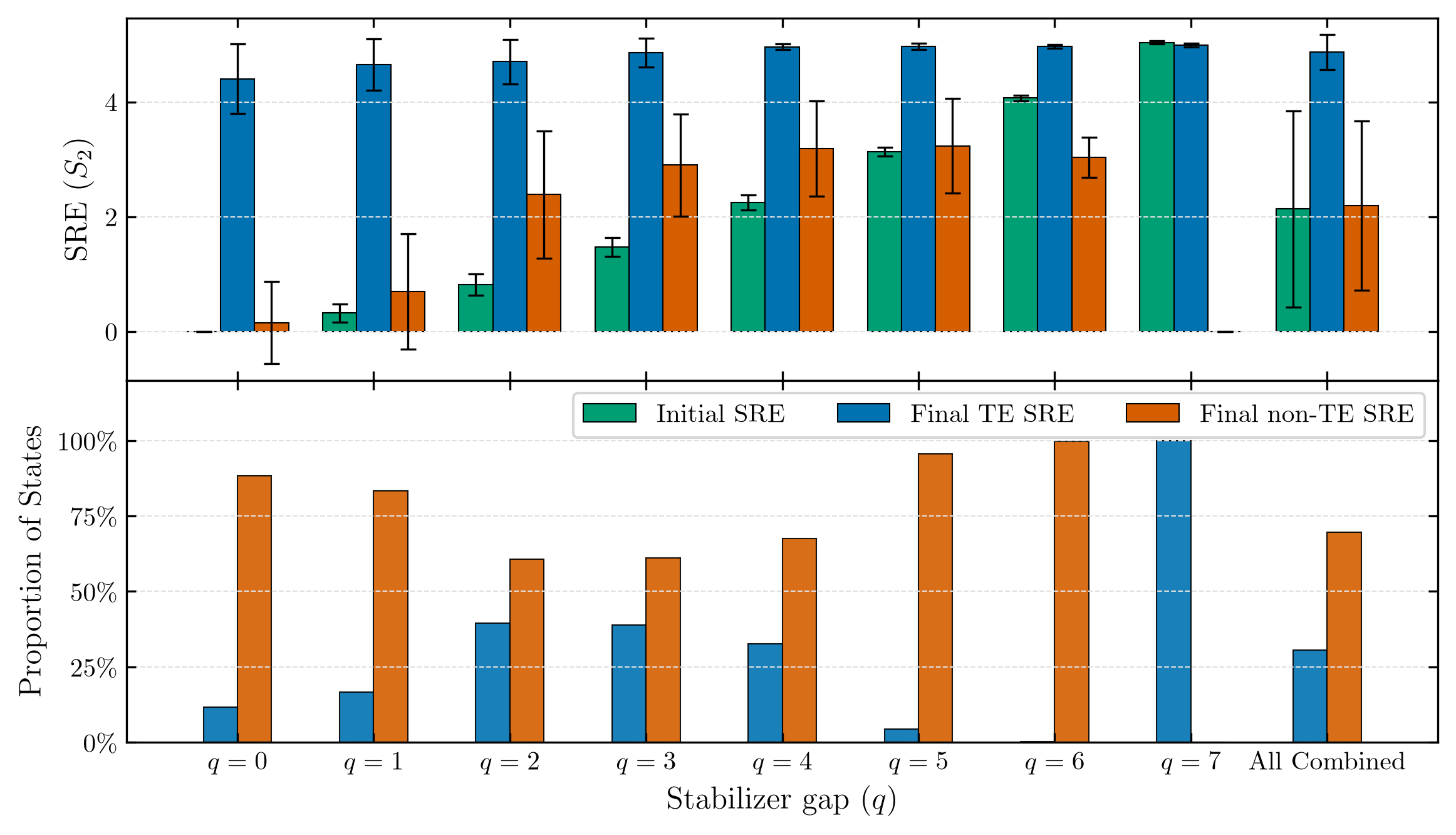}
		\caption{Top: Comparison of SRE of the initial $2000$ seed $7$-qubit states and the states obtained after the $2500$-step gradient descent optimization separated by TE-ness of the final states. Bottom: Proportion of states that ended up being TES or Non-TES  }
		\label{fig:7q_2000_Hist}
		\vspace{0.5cm} 

		\includegraphics[width=0.90\textwidth]{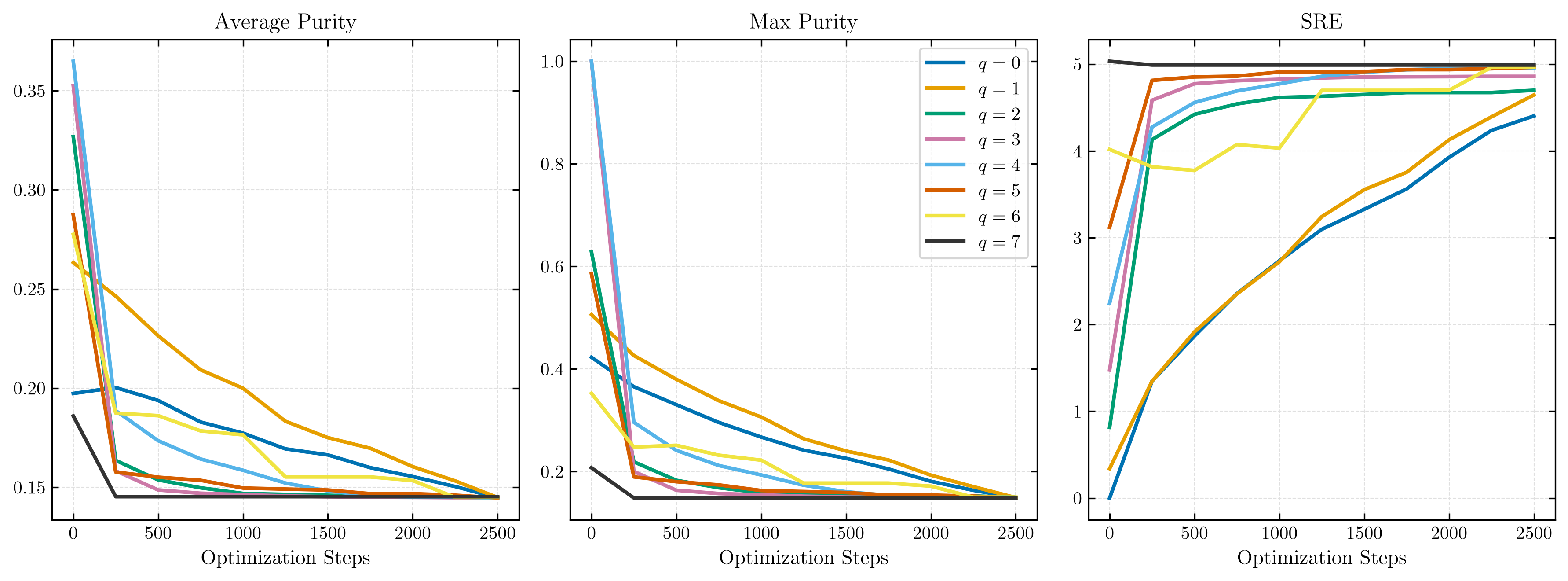}
		\caption{Average Purity, Maximal Purity and SRE for $7$-qubit seed-states that remain TES after $2000$ steps of gradient descent optimization.}
		\label{fig:7q_2000_TE}
		\vspace{0.5cm}
		\includegraphics[width=0.90\textwidth]{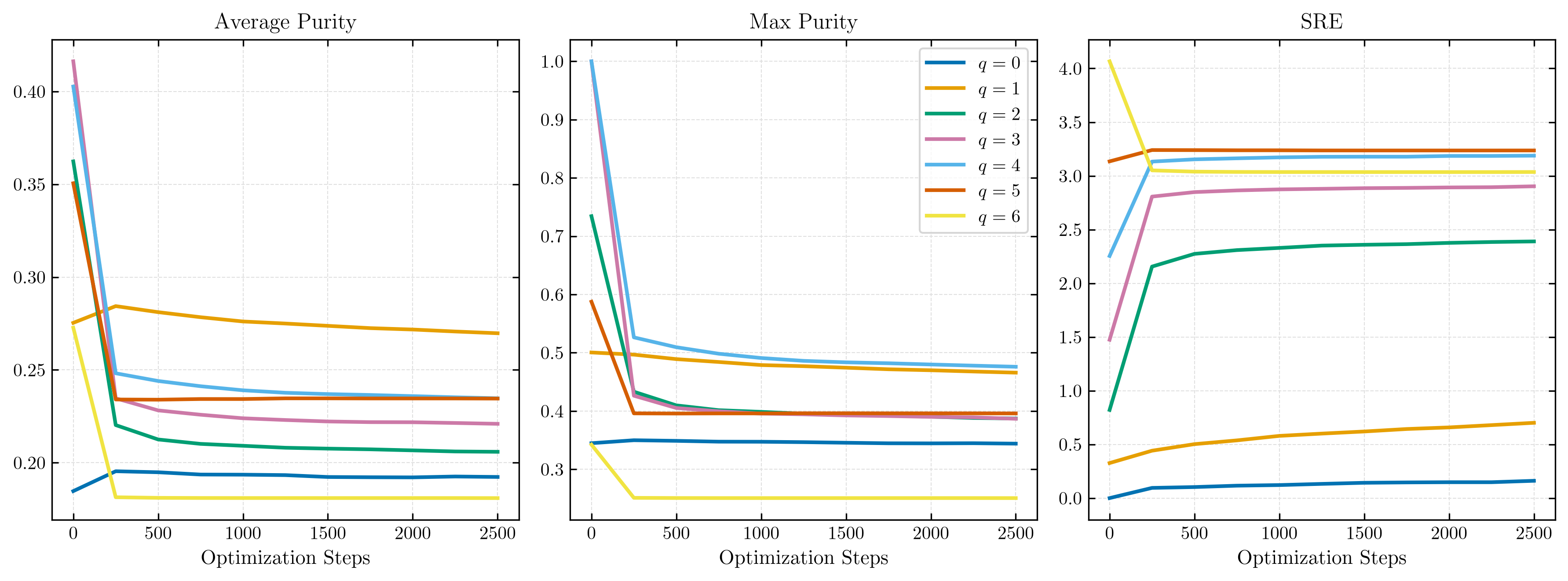}
		\caption{Average Purity, Maximal Purity and SRE for $7$-qubit seed-states that remain non-TES after $2000$ steps of gradient descent optimization.}
		\label{fig:7q_2000_NTE}
	\end{figure}
\end{widetext}

\end{document}